# Recent Advances on $\pi$-Conjugated Polymers as Active Elements in High Performance Organic Field-Effect Transistors

Lixing Luo[1], Wanning Huang[1], Canglei Yang[1], Jing Zhang[1,*], Qichun Zhang[2,*]

[1] Mr. L. Luo, Mrs. W. Huang, Mr. C. Yang, Prof. Dr. J. Zhang
Key Laboratory for Organic Electronics and Information Displays & Institute of Advanced Materials, Jiangsu National Synergetic Innovation Center for Advanced Materials
Nanjing University of Posts & Telecommunications
9 Wenyuan Road, Nanjing 210023, China
E-mail: iamjingzhang@njupt.edu.cn

[2] Prof. Dr. Q. Zhang
Department of Materials Science and Engineering
City University of Hong Kong
Kowloon, Hong Kong SAR, 999077, China
Email: qiczhang@cityu.edu.hk

**Abstract:** As high-performance organic semiconductors, $\pi$-conjugated polymers have attracted much attention due to their charming advantages including low-cost, solution processability, mechanical flexibility, and tunable optoelectronic properties. During the past several decades, the great advances have been made in polymers-based OFETs with p-type, n-type or even ambipolar characterics. Through chemical modification and alignment optimization, lots of conjugated polymers exhibited superior mobilities, and some mobilities are even larger than 10 $cm^2\ V^{-1}\ s^{-1}$ in OFETs, which makes them very promising for the applications in organic electronic devices. This review describes the recent progress of the high performance polymers used in OFETs from the aspects of molecular design and assembly strategy. Furthermore, the current challenges and outlook in the design and development of conjugated polymers are also mentioned.

## 1. Introduction

Over the past three decades, conjugated polymers, as promising electroactive components，have many potential applications in low-temperature solution processing and low-cost high-throughput optoelectronic devices, such as organic field-effect transistors (OFETs) [1-5], electrochromic devices [6-8], memory [9, 10], organic solar cells (OSC) [11-15] and organic light-emitting diodes (OLED) [16, 17]. With the joint efforts from molecular design, processing technology and device construction, superior electronic properties have been achieved. Usually, the main mechanisms of charge transport in conjugated polymers can be separated into intrachain and interchain transport modes [18-20]. Since carriers are much easy to transport through the covalent bonds, the predicted mobility of intrachain transport is very high. However, since the disordered accumulation of the conjugated main chain and the defects in the film will inhibit the transport of charges, the carriers tend to be transported along the $\pi$-$\pi$ stacking direction, which is commonly seen in highly-crystalline or long-range ordered semicrystalline films. In addition, the long and conjugated skeleton with good planarity can connect the short-range ordered regions of the film, thereby establishing a carrier transport channel in the macroscopically disordered polymer film [21, 22]. Because of the complicated morphology, polymers usually possess more electron traps than small molecule semiconductors [23]. Nevertheless, in recent years, the mobility of many conjugated polymers has surpassed 10 $cm^2$ $V^{-1}$ $s^{-1}$ through molecular design strategies (for example, side chain engineering, chemical substitution, and weak donor-strong acceptor systems) [24].

For conjugated polymers, adjusting the electron-donating or electron-withdrawing strength of the repeating units can alter the energy level of the front orbital levels (the highest occupied molecular orbital (HOMO) and the lowest unoccupied molecular orbital (LUMO)), which can lead to the effective charge injection and extraction from the electrodes [25]. The typical electron-deficient acceptors of OFETs include diketopyrrolopyrrole (DPP), isoindigo (IID), naphthalimide (NDI), benzodifurandione-based oligo(*p*-phenylene vinylene) (BDOPV) and so on. Currently, most high-performance OFETs based on D-A polymers are p-type and the number of n-type devices is small, mostly because the electron-withdrawing building units with low LUMO energy levels and high electron transport channel are rare. Therefore, the research on ambipolar or n-type polymers is becoming more and more important in the field of high performance OFETs to realize the CMOS function. An effective way to achieve ambipolar or n-type polymers is to reduce their LUMO energy level to avoid the damage from moist and oxygen.

Through the special processing procedure, the polymer chains can be accumulated in an orderly manner to depress defects and greatly improve the carrier mobility of the polymer materials. In order to realize low-cost large-area semiconductor devices, efficient and easy-to-operate manufacturing methods are essential. In polymer film fabrication, scientists try to increase the crystallinity of thin-films, by controlling the evaporation rate [26, 27] or using external physical force [28-30]. The application of external stress (shearing force, centrifugal force or capillary force) contributes to the formation of directional structures in spin-coating, drop-casting, bar-coating or solution shear process. In single crystal and nanowire growth, the preparation of intra-chain pathway dominating polymer structure can help us obtain high intrinsic mobility of the material. From the early poly(alkylthiophene)s (P3ATs) to the now eye-catching donor-acceptor (D-A) polymers, most single crystals/nanowires are grown by solution-assembly methods, including solvent-assisted crystallization [31-33], solution annealing [34], self-seeding [35, 36], etc.

In this review, we describe recent advances from three aspects: (1) The charge transport mechanism in the conjugated polymer structure; (2) Polymer materials of high-performance p-, n- or ambipolar transporting; and (3) Various methods used for aligned polymer structure growth.

## 2. Molecular stacking model of conjugated polymers

Because the backbone chain segments are easy to interpenetrate and entangle with each other, polymers need to cross a large free energy barrier to self-assemble into single crystals or high crystallinity nanostructures [37]. Considering that the ordering in organic semiconductor materials is a key factor in the performance of the device, optimizing the aggregated structure of the polymer semiconductor is one of the priority tasks to maximize the mobility of carriers [38]. From the perspective of basic research, the regulation of the aggregated structure of conjugated polymer molecules is helpful for us to understand the process of polymer crystallization and the transport mechanism of carriers in the polymer.

The different crystalline orientations are referred as edge-on, face-on, and chain-on (Fig. 1a), corresponding to the planar conjugated backbones lying perpendicular/parallel or standing perpendicular to the substrate [39]. Most previous studies reveal that polymer systems have three different transmission paths. (1) Along the direction of the polymer backbone ($I_1$ represents the direction of intrachain transport), the delocalized π orbitals on a complete chain extend to the entire polymer chain due to conjugation, so that the charge can be rapidly transferred throughout the molecule. If polymer chains with high planarity, high purity, and defect-free can be realized, the high charge mobility can be achieved along the direction of the conjugated backbone chain when the conjugated length increases. For example, the intrachain mobility of the stepped polymer poly(p-phenylenes) is up to 600 $cm^2\ V^{-1}\ s^{-1}$, three orders of magnitude higher than its interchain mobility [40]. (2) Along the $\pi$-$\pi$ stacking direction of the conjugated main chain ($I_2$ represents the direction of interchain transport), most of the charges are actually transferred through the hopping between adjacent π orbitals in the polymer film system. Therefore, a short $\pi$-$\pi$ stacking distance, or compact face-to-face $\pi$ stacking, is an important indicator that is essential to obtain higher mobility. (3) The charge transfer in the chain-to-chain direction is mostly restricted, and the carrier transmission efficiency is lowest because of the insulating effect of the long alkyl chain and the steric hindrance of the side chain substituents, which impede the effective transport of carriers. In short, the charge mobility increases with the augment of film crystallinity and the shortening of the π-π stacking distance between the main chains, where long-range order and close $\pi$-$\pi$ stacking often mean high mobility [41]. In order to obtain thin films with high crystallinity, many technologies have been developed to increase the ordered region areas of the film, reduce interface defects, and constrain polymer chain orientation to establish an effective channel for charge transfer along the $\pi$-$\pi$ stacking direction.

Since long-range ordered *π*-*π* stacking is not a necessary limitation for good carrier transport, many polymer films with low crystallinity still exhibit excellent transport properties. Generally speaking, different films can be divided into several types, namely, semicrystalline films (Fig. 1b), disordered aggregates (Fig. 1c), and completely amorphous films (Fig. 1d) [21]. Among them, the semicrystalline films have good crystallinity, so that the close *π*-*π* stacking reduces the barrier of the charge hopping between adjacent *π* orbital units, resulting in the high mobility of the films, such as poly(3-hexylthiophene-2,5-diyl) (P3HT) [42] and poly[2,5-bis(3-alkylthiophen-2-yl)thieno(3,2-b)thiophene] (PBTTT) [43] thin-films. Disordered aggregates present relatively poor crystallinity, but sometimes associating with good device performance, for example, high-performance D-A polymer diketopyrrolopyrrole-dithienylthieno[3,2-b]thiophene (DPP-DTT) [44] and poly[N,N′-bis(2-octyldodecyl)naphthalene-1,4,5,8-bis(dicarboximide)-2,6-diyl]-alt-5,5′-(2,2′-bithiophene) (P(NDI2OD-T2)) [45]. Long-range ordered *π*-*π* stacking is not a necessary limitation for good carrier transport, because smaller domains in short-range ordering of a few repeating units are connected by rigid polymer chains (highlighted in red), allowing carriers to be transported along the conjugated chain efficiently and quickly. Although this thin film with many amorphous regions lacks sufficient long-range order, charge transport in these polymers occurs predominantly along the conjugated polymer backbone and requires only occasional intermolecular hopping via interchain *π*-*π* stacking. The efficient intrachain transport of carriers along the rigid polymer backbone and the interchain transport in the tightly packed aggregates synergetically establish a "highway" for carrier transport in films. The same principle is also applicable to semicrystalline films. By inserting alkyl chain into the molecular backbone of DPP-C0 (poly-[2,5-bis(4-decyltetradecyl)-3-(5''-methyl-[2,2':5',2''-terthiophen]-5-yl)-6-(5-methylthiophen-2-yl)-2,5-dihydropyrrolo[3,4-c]pyrrole-1,4-dione]) for blocking the charge-carrier transport along the polymer chain, highly soluble DPP-C3 was obtained [22]. When there only exists charge transfer along *π*-*π* direction, the mobility of the thin film device is only 0.015 $cm^2 V^{-1} s^{-1}$. However, after blending it with 1 wt% DPP-C0 that has good intrachain transport ability, the film mobility increases to 1.14 $cm^2 V^{-1} s^{-1}$. The efficient intrachain transport of carriers along the rigid polymer backbone and the interchain transport in the tightly packed aggregates synegetically establish a "highway" for carrier transport in films, which is diffeicultly achieved in the completely-amorphous films without any regularity. Certainly the OFET results are also dependent on many factors, such as film thickness, electrode materials, dielectric layers, and the contacts between the electrode/semiconductor and the dielectric/semiconductor. Generally, gold electrodes and octadecyltrichlorosilane(OTS)-modified $SiO_2$/Si substrates are the most common choices to fabricate most OFETs devices in this review if there is no special mention.

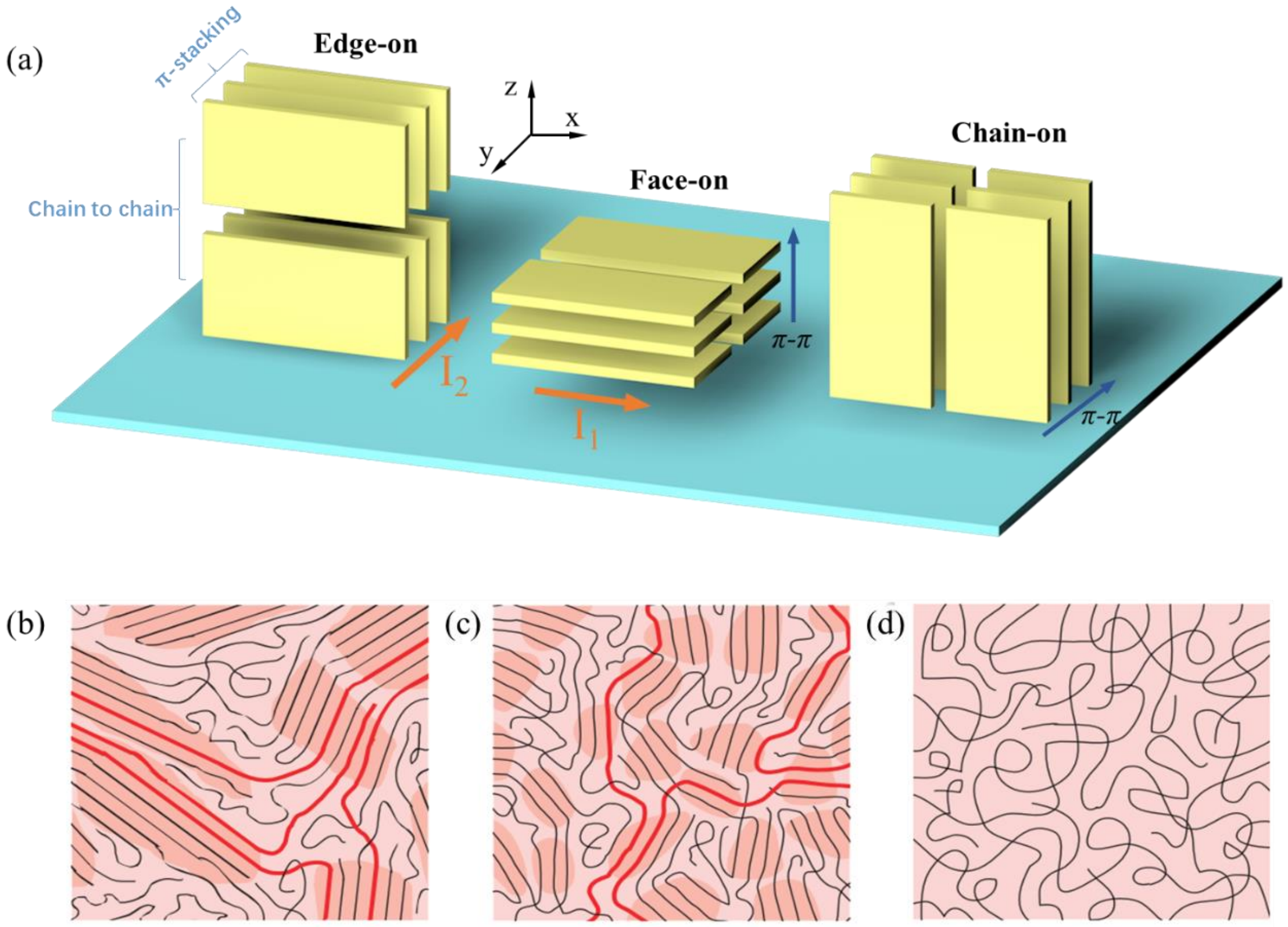

**Fig. 1** (a) Simple models of chain orientation in crystallites. From left to right: Edge-on, face-on, and chain-on lamellae with $\pi$-$\pi$ stacking on the plane of the film. Schematic diagram of (b) semicrystalline films, (c) disordered aggregates, and (d) completely amorphous films. Reproduced from Ref. [21].

# 3. High-Performance $\pi$-Conjugated Polymers

## 3.1 P-Type Conjugated Polymers

The p-type semiconducting polymers have been developed rapidly due to their flexibility, good hole mobility, and excellent device stability [2, 46]. The HOMO energy levels of these polymers range from -5.0 to -5.5 eV for efficient hole collection/injection from/into the electrodes [47]. Till now, p-type polymer-based field-effect transistors can be optimized through the molecular design and the tuning of fabricating technology to display the maximum hole mobility ($\mu_h$) up to ~92 cm$^2$ V$^{-1}$ s$^{-1}$. The following part summarizes the latest research progress of high-performance p-type polymers based on DPP, indenodithiophene (IDT), cyclopentadithiophene (CDT), and IID units, and the related structural formulas mentioned in this part are listed in Fig. 2.

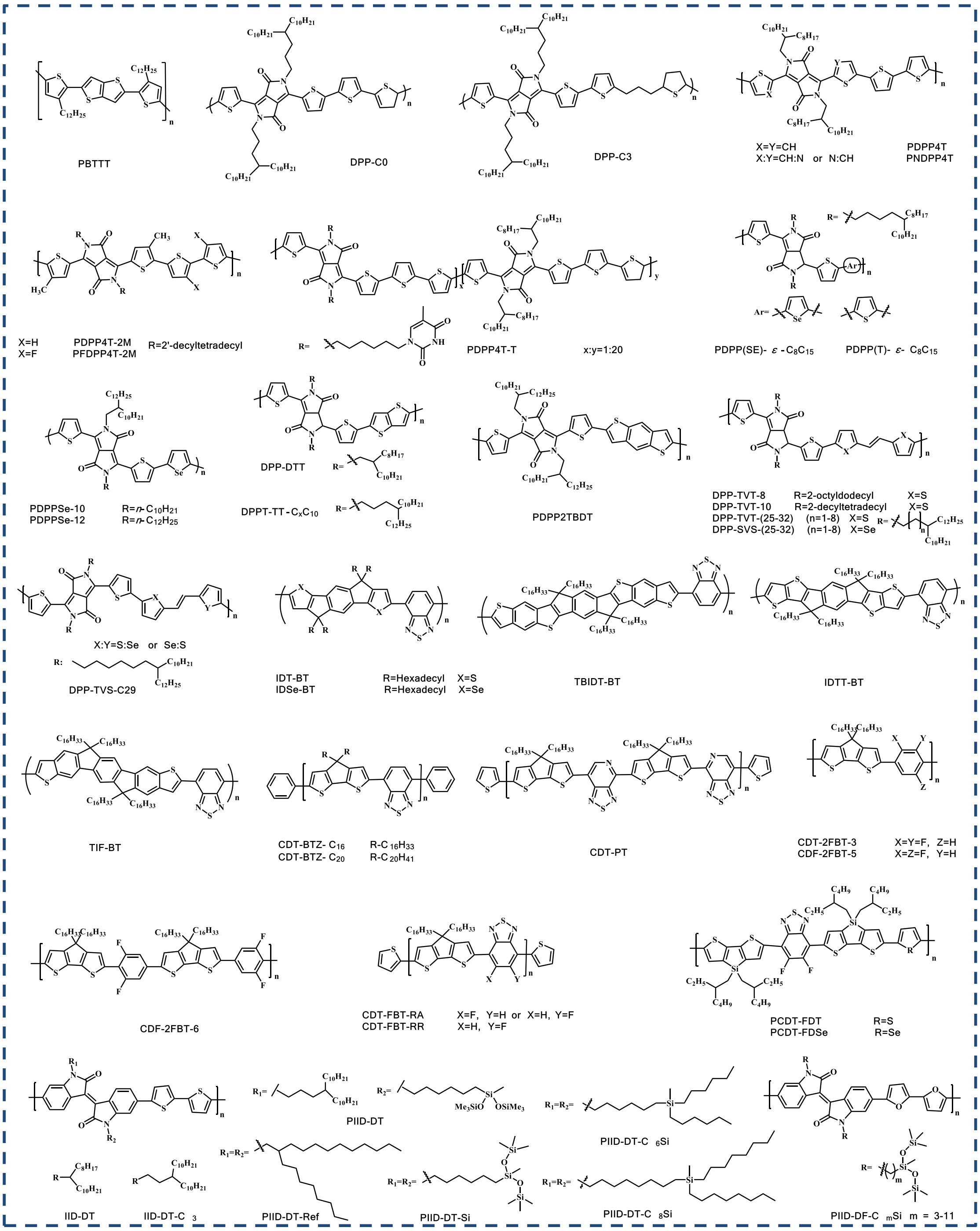

**Fig. 2** DPP-, IDT-, CDT- and IID-based high-performance polymers.

### 3.1.1 DPP-based polymers

Since diketopyrrolopyrrole (DPP) polymer was first applied in OFETs [48], the DPP motif has been widely introduced into organic polymer backbones due to its advantageous properties (Table 1) including high planarity and polar nature [49]. For example, diketopyrrolopyrrole–quaterthiophene copolymer (PDPP4T) containing bithiophene (T2) group, prepared via microwave-assisted Stille coupling, revealed an edge-on orientation manner in the spin-coating thin-films with the $\mu_h$ value of 0.74 cm$^2$ V$^{-1}$ s$^{-1}$, while the higher number-average molecular mass ($M$n, 47 kDa) of PDPP4T, prepared through Yamamoto coupling polymerization, displayed a much higher $\mu_h$ up to 5.50 cm$^2$ V$^{-1}$ s$^{-1}$ [50]. Moreover, a new DPP polymer PNDPP4T with thienyl and thiazoyl units as linkers was designed and synthesized, and its spin-coating thin-film demonstrated a high crystallinity and a good hole mobility of 3.05 cm$^2$ V$^{-1}$ s$^{-1}$ [51]. Through cyclic voltammetry (CV) measurement and theoretical calculation, the polymer is believed to own deep-lying energy levels, which is supposed to result in superior performance. Similarly, after replacing thiophene with methylated thiophene unit, the low crystalline PDPP4T-2M still exhibited a high hole mobility of 11.16 cm$^2$ V$^{-1}$ s$^{-1}$, which could be attributed to good aggregations even in the disordered area of the thin films [52]. However, the fluorinated polymer PFDPP4T-2M based thin-film transistor illustrated a relatively low hole mobility of 1.80 cm$^2$ V$^{-1}$ s$^{-1}$, due to the poor crystallinity and the enhanced face-on orientation of the thin-films. Apart from the polymer backbone tuning, the self-assembly and carrier mobility can further be engineered by the terminated group modification. For instance, the thymine substitution at the end of the alkyl side chain of PDPP4T to generate polymer PDPP4T-T could enable H-bonding among thymine groups to enhance the chain-to-chain interactions and promote the orderly of the polymer chain [53]. Compared with the original PDPP4T with alkyl chains, the PDPP4T-T based thin-film transistor shows a higher hole mobility of 9.1 cm$^2$ V$^{-1}$ s$^{-1}$.

Table 1. Molecular weight, optical and electrochemical properties, OFET device structure, and performance of DPP-based high-performance p-type polymers.

| Polymer | *M*n[kDa] | $E_g^{opt}$[eV] | HOMO[eV] | LUMO[eV] | Device structure | $\mu_h$,max [cm$^2$ V$^{-1}$ s$^{-1}$] | Refs |
|---|---|---|---|---|---|---|---|
| PDPP4T | 47 | 1.30 | -5.30 | -4.00 | BGBC | 5.50 | 50 |
| PNDPP4T | 75.4 | 1.45 | -5.48 | -4.03 | BGBC | 3.05 | 51 |
| PDPP4T-2M | 40.4 | 1.40 | -5.43 | -4.03 | BGBC | 11.16 | 52 |
| PFDPP4T-2M | 116.7 | 1.42 | -5.59 | -4.17 | BGBC | 1.80 | 52 |
| PDPP4T-T | 99.5 | 1.40 | -5.36 | -3.64 | BGBC | 9.10 | 53 |
| PDPPSE-$\varepsilon$-$C_8C_{15}$ | 38.2 | 1.26 | -4.94 | -3.68 | BGTC | 12.25 | 54 |
| PDPPT-$\varepsilon$-$C_8C_{15}$ | 50.8 | 1.28 | -4.95 | -3.67 | BGTC | 8.32 | 54 |
| PDPPSe-10 | 55.1 | 1.27 | -5.20 | -3.64 | BGBC | 8.10 | 55 |
| PDPPSe-12 | 60.7 | 1.26 | -5.20 | -3.66 | BGBC | 9.40 | 55 |
| DPP-DTT | 110 | 1.70 | -5.20 | -3.50 | BGBC | 10.50 | 56 |
| DPP-DTT | 55 | — | — | — | BGTC | 8.25 | 57 |
| DPP-DTT | — | — | — | — | BGTC | 5.00 | 44 |
| DPPT-TT-$C_{12}C_{10}$ | 51 | 1.30 | -5.20 | -3.90 | BGTC | 3.70 | 58 |
| DPP-DTT(SEBS) | — | — | — | — | BGBC | 1.35 | 68 |
| PDPP2TBDT | — | — | — | -3.8 | BGTC | 7.42 | 59 |
| DPP-TVT-8 | 70 | — | -5.30 | — | BGBC | 4.50 | 61 |
| DPP-TVT-10 | 73.5 | — | -5.28 | — | BGBC | 8.20 | 61 |
| DPP-TVT-29 | 519 | 1.28 | -5.25 | -3.97 | BGTC | 8.74 | 62 |
| DPP-SVS-29 | 34.1 | 1.23 | -5.14 | -3.91 | BGTC | 17.80 | 63 |
| DPP-TVS-29 | 74.5 | 1.24 | -5.17 | -3.93 | BGTC | 8.20 | 65 |

Solution processable polymers PDPP(SE)-$\varepsilon$-$C_8C_{15}$ and PDPP(T)-$\varepsilon$-$C_8C_{15}$, composed of a DPP skeleton and finely-tuned branched side chains ($\varepsilon$-branched), have been demonatrated to show high hole mobilities of 12.25 and 8.32 cm$^2$ V$^{-1}$ s$^{-1}$, respectively [54]. Although similar layered stacking structures and small spacing differences were found in their thin-films compared to PDPP(T)-$\varepsilon$-$C_8C_{15}$, larger fibrils in PDPP(SE)-$\varepsilon$-$C_8C_{15}$ film facilitated a better charge transport capability [55]. By contrast, PDPPSe-10 and PDPPSe-12 polymers via replacing branching alkyl chain with a linear one possessed a more planar conjugated skeleton due to the reduced steric hindrance. The grazing incidence wide-angle X-ray scattering (GIWAXS) data (Fig. 3) unveiled that the polymer chains preferred to the edge-on orientation with obvious better crystallinity than that of face-on stacked PDPPSE films. The hole mobilities of PDPPSe-10 and PDPPSe-12 thin films, measured in air condition, were 8.1 cm$^2$ V$^{-1}$ s$^{-1}$ and 9.4 cm$^2$ V$^{-1}$ s$^{-1}$, respectively, 6-7 times higher than that of PDPPSE (1.35 cm$^2$ V$^{-1}$ s$^{-1}$).

Employing another donor fragment thieno[3,2-b]thiophene (TT) as the building unit, the new synthesized D-A conjugated polymer DPP-DTT under high-Mn value exhibited high hole mobility (up to 10.5 cm$^2$ V$^{-1}$ s$^{-1}$) and long-term stability under ambient condition [56]. Furthermore, by dissolving DPP-DTT and polystyrene simultaneously in dichlorobenzene, the low-molecular-weight low-mobility DPP-DTT could be converted into high crystallinity and high-performance state, from which the spin-coating thin-film device mobility reached to 8.25 $cm^2 V^{-1} s^{-1}$ [57]. The polystyrene in the blend was conducive to the macromolecular self-assembly and crystallization for the growth of long-range high crystalline orders by thermal treatment above the glass transition temperature ($T_g$) of polystyrene. All examples mentioned above require a high-temperature postdeposition treatment for the semiconductor film annealing, but some exceptions like incompatible hydrocarbon binders or waxes incorporating into the polymer DPP-DTT channel can drive rapid phase separation and polymer molecular ordering at room temperature to obtain high OFET mobility of about 5 $cm^2 V^{-1} s^{-1}$ with excellent thermal stability [44]. In order to study the relationship between the branched side chain geometry and the charge carrier mobility in DPP-based polymers, DPP-DTT polymers with different branched alkyl side chains were synthesized [58]. A high hole mobility of 3.7 $cm^2 V^{-1} s^{-1}$ was achieved from the blending films of edge-on (DPPT-TT-$C_6C_{10}$) and face-on (DPPT-TT-$C_{14}C_{10}$) orientation polymers, probably due to the formation of three-dimensional (3D) conductive paths. Using benzodithiophene (BDT) as the donor unit, diketopyrrolopyrrole-thiophene-benzodithiophene (PDPP2TBDT) based thin-film transistor illustrated a relative low hole mobility of 0.72 $cm^2 V^{-1} s^{-1}$, while the 1D polymer nanowires, prepared by the drop-casting method, showed a high hole mobility of 7.42 $cm^2 V^{-1} s^{-1}$ for the reduced structural and energetic disorder [59].

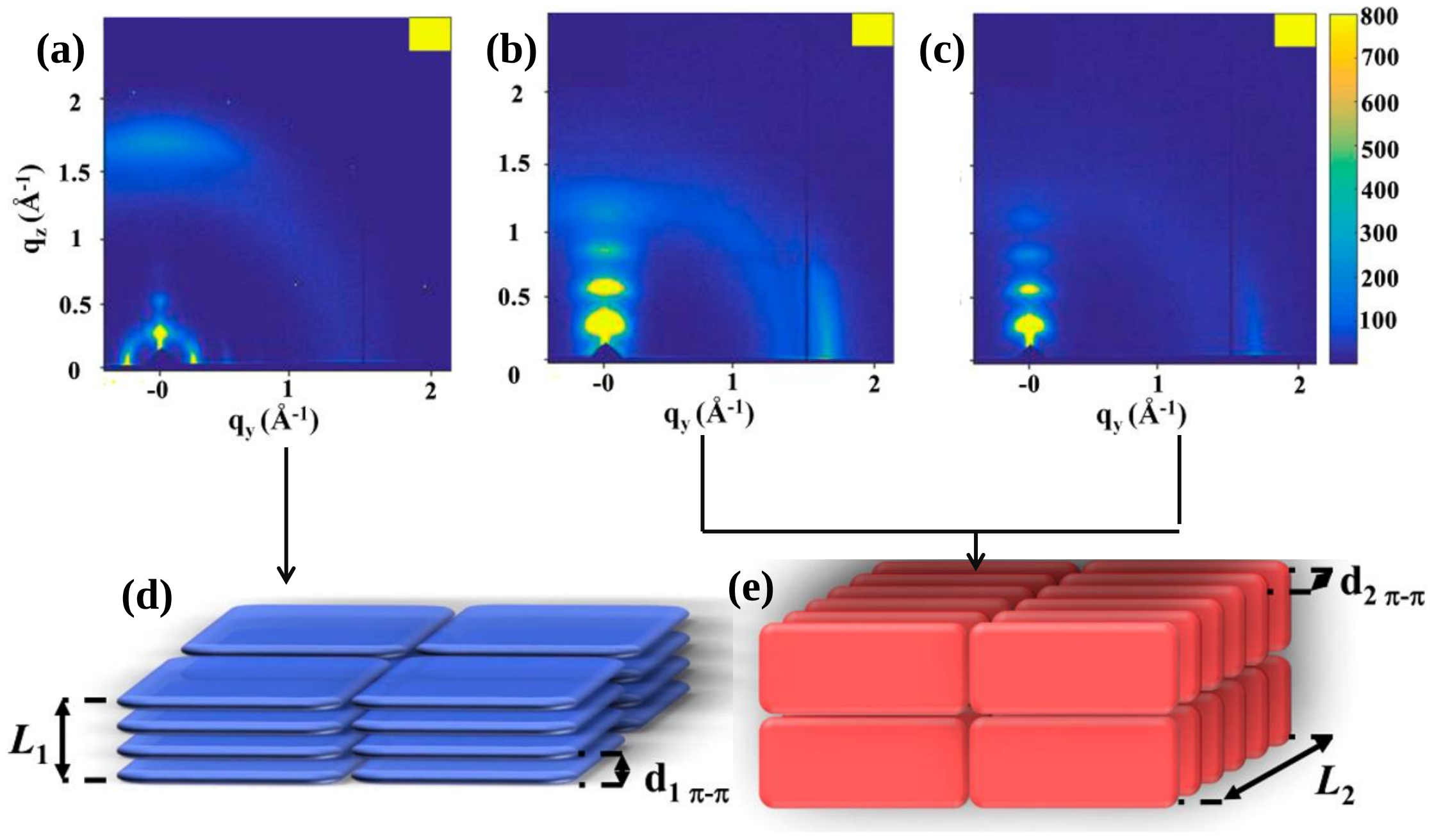

**Fig. 3** GIWAXS patterns of PDPPSe (a), PDPPSe-10 (b), and PDPPSe-12 (c) deposited on OTS-modified $SiO_2$/Si substrates after thermal annealing at 180 °C. Schematic illustration for stacking orientation of PDPPSe (d, face-on) and PDPPSe-10, PDPPSe-12 (e, edge-on). Reproduced from Ref. [55].

The conjugated polymers have been developed as softer semiconductors with high charge carrier mobilities rivaling that of poly-Si, however, the lower stretchability performance limits their mechanical ductility and high carrier mobility at large strains. To address this issue, DPP-DTT polymer was mixed with the elastomer matrix polystyrene-block-poly(ethylene-butylene)-block-polystyrene (SEBS) to substantially improve the stretchability of polymer semiconductors, without affecting charge transport mobility [60]. The nanometer-scale polymer nanofibrils, formed in elastomer, reduced the mechanical modulus and glass transition temperature and elevated the mechanical ductility. Thus, the deformable elastomer and good charge transfer can be maintained (1.35 $cm^2$ $V^{-1}$ $s^{-1}$ under 100% strain) while their interfacing with the deformable elastomer prevented crack propagation (Fig. 4).

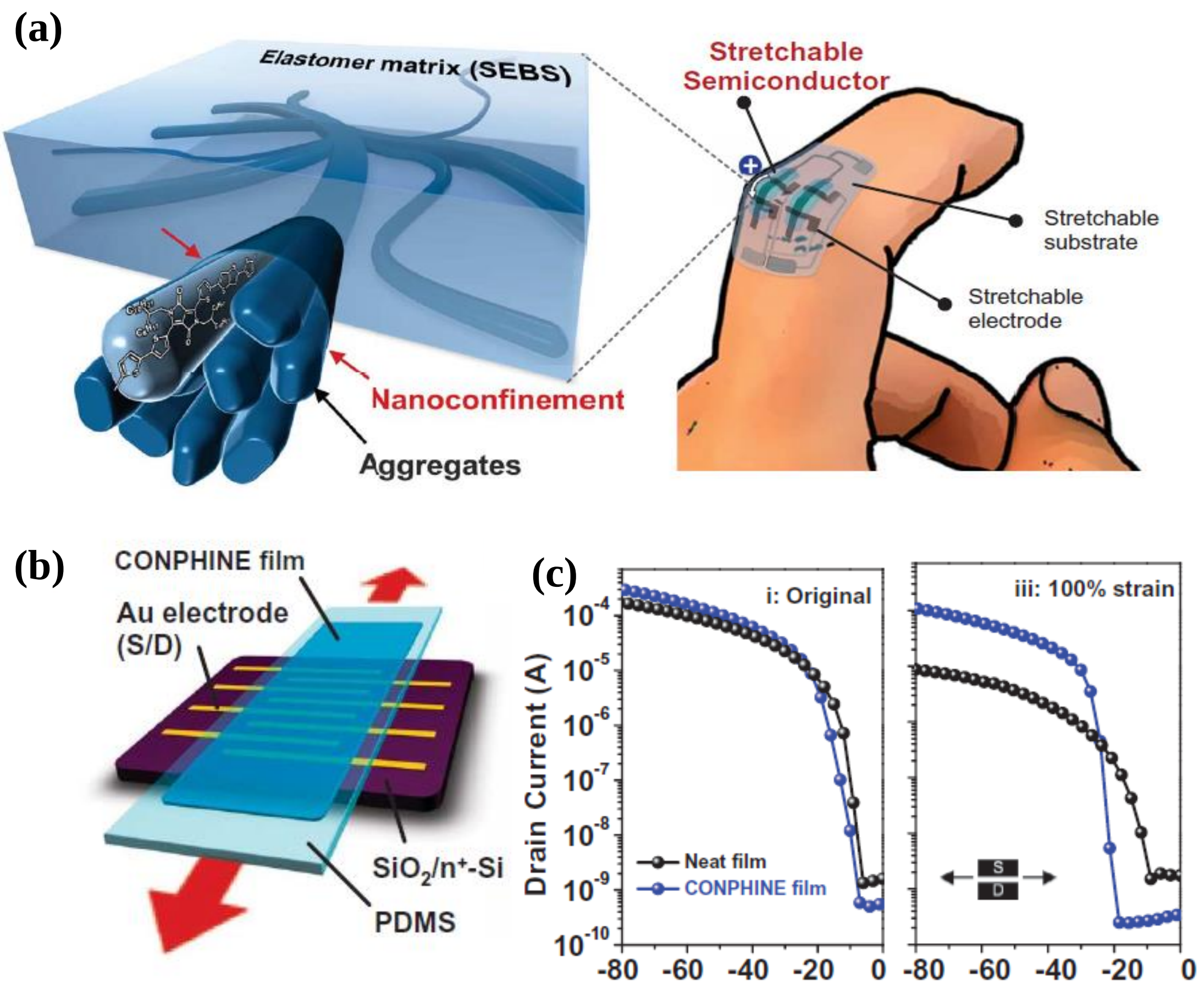

**Fig. 4** (a) Schematic 3D illustration of the morphology for confinement film. (b) Schematic of the soft contact lamination method for characterizing the electrical performance of a semiconducting layer upon stretching. (c) Transfer curves obtained from the confinement film and the Neat film in its original condition and under 100% strain perpendicular to the charge transport direction. Reproduced from Ref. [60].

Furthermore, a series of diketopyrrolopyrrole-thiophene-vinylene-thiophene (DPP-TVT) polymers were reported, where the D-A structure allowed the effective intramolecular charge transfer that promoted intermolecular interactions, and the incorporation of TVT units enhanced the coplanarity of the conjugated $\pi$-backbones for much close stacking [61]. The hole mobilities of copolymers DPP-TVT-8 and DPP-TVT-10 devices are calculated to be 4.5 $cm^2 V^{-1} s^{-1}$ and 8.2 $cm^2 V^{-1} s^{-1}$, respectively. Then, polymers (DPP-TVT-25 to DPP-TVT-32) with branched alkyl groups and various linear spacer groups ($C_2$ to $C_9$) were synthesized [62]. Grazing incidence X-ray diffraction analysis proved that polymers with even-numbered linear spacer groups had shorter d-spacing values than those with odd-numbered linear spacer groups, indicating more closely molecular packing and stronger intermolecular interactions. However, the odd-even effect for linear alkyl spacers greater than $C_7$ no longer existed. Therefore, the highest hole mobility (8.74 $cm^2 V^{-1} s^{-1}$) was realized from polymer DPP-TVT thin-film with the $C_6$ spacer unit by solution-shearing approach. Moreover, when S atoms in the TVT unit was substituted by Se atoms, the diketopyrrolopyrrole-selenophene-vinylene-selenophene (DPP-SVS) polymer with the $C_6$ spacer via the same Stille coupling reaction showed a maximum mobility of 17.8 $cm^2 V^{-1} s^{-1}$ [63, 64]. For partial Se replacement, the asymmetric thiophene-ethylene-selenothiophene (TVS) donor unit and DPP fragment were polymerized into the conjugated polymer DPP-TVS-C29, and the as-prepared thin-film OFET devices using non-chlorinated solvents exhibited a hole mobility up to 8.2 $cm^2 V^{-1} s^{-1}$ [65].

### 3.1.2 IDT-based polymers

Considering the polymer structure design, polymers composing of rigid indenodithiophene (IDT) units usually presented excellent solution-processable ability due to the bridging $sp^3$ carbon atom [66]. The substituents on the bridging carbon, usually aliphatic or aromatic groups, can enhance the solubility and play an important role in the thin-film microstructure [67]. The near-amorphous indacenodithiophene-co-benzothiadiazole (IDT-BT) polymer thin-film still provide a hole mobility as high as 1 $cm^2 V^{-1} s^{-1}$ (Table 2) [68]. In addition, by increasing the Mn from 38 kDa to ~80 kDa, the $\mu_h$ value of IDT-BT polymer film was further enhanced to 3.6 $cm^2 V^{-1} s^{-1}$, because of the better crystallinity, better domain connectivity, and smaller grain boundaries [69]. Besides, through the substitution of selenium atoms, the as-prepared copolymer IDSe-BT possessed a highly coplanar skeleton under the enhanced intermolecular interaction of the conjugated backbones with a hole mobility up to 3.2 $cm^2 V^{-1} s^{-1}$. After introducing a highly effective electron-blocking CuSCN layer, its hole mobility can reach to a maximum of 6.4 $cm^2 V^{-1} s^{-1}$ [70]. When thieno[b]indenodithiophene (TBIDT) as an extended IDT moiety was brought into the copolymer backbone (TBIDT-BT), an average saturation hole mobility of 0.9 $cm^2 V^{-1} s^{-1}$ was achieved, which was lower than that of IDT-BT (1.5 $cm^2 V^{-1} s^{-1}$) [71]. The reduced electrostatic stabilizing interactions between the peripheral thiophene of the fused core and the BT unit resulted in a poorer stability of the planar skeleton geometry, which influenced the charge carrier mobility. Therefore, in order to obtain a higher frame flatness, the polymer indenodithieno[3,2-b]thiophene-benzothiadiazole (IDTT-BT) containing IDTT and BT units has been prepared to show a good solubility and a high framework planarity.

Its thin-film device presented a mobility of 6.6 cm$^2$ V$^{-1}$ s$^{-1}$ after annealing at 270°C [72]. Further optimized device with a thin CuSCN layer (blocking electron injection and facilitating hole injection) showed a promoting hole mobility of 8.7 cm$^2$ V$^{-1}$ s$^{-1}$. Furthermore, the researchers extended the IDT aromatic core into the dithiopheneindenofluorene (TIF) repeat unit, prepared through an intramolecular palladium catalyzed C-H cyclization strategy. The resulting TIF-BT copolymer has a two-fold increase in hole mobility, reaching the value of almost 3 cm$^2$ V$^{-1}$ s$^{-1}$ in bottom-gate top-contact organic field-effect transistors [73].

Table 2. Molecular weight, optical and electrochemical properties, OFET device structure, and performance of IDT-, CDT- and IID-based, high-performance p-type polymers.

| Polymer | $M$n[kDa] | $E_g^{opt}$[eV] | HOMO[eV] | LUMO[eV] | Device structure | $\mu_h$,max [cm$^2$ V$^{-1}$ s$^{-1}$] | Refs |
|---|---|---|---|---|---|---|---|
| IDT-BT | 38 | 1.70 | -5.40 | -3.70 | TGBC | 1.20 | 68 |
| IDT-BT | 80 | — | — | — | TGBC | 3.60 | 69 |
| IDSe-BT | 119 | 1.59 | -5.22 | -3.63 | TGBC | 6.40 | 70 |
| TBIDT-BT | 62.1 | 1.90 | -5.50 | -3.60 | TGBC | 0.90 | 71 |
| IDTT-BT | 76 | 1.71 | -5.40 | -3.69 | TGTC | 8.70 | 72 |
| TIF-BT | 56.5 | 2.00 | -5.70 | -2.70 | TGBC | 2.80 | 73 |
| CDT-BTZ-$C_{16}$ | 10.2 | — | — | — | TGBC | 0.17 | 74 |
| CDT-BTZ-$C_{16}$ | 35 | — | — | — | BGBC | 3.30 | 75 |
| CDT-BTZ-$C_{16}$ | 50 | — | — | — | BGTC | 5.50 | 76 |
| CDT-BTZ-$C_{20}$ | 30 | — | — | — | BGBC | 11.40 | 77 |
| CDT-PT | 140 | — | — | — | BGBC | 47.00 | 78 |
| CDT-PT CDT- | 50 | — | — | — | BGTC | 92.64 | 79 |
| 2FBT-3 | 34 | 2.05 | -5.30 | -3.25 | BGBC | 2.80 | 80 |
| CDT-2FBT-5 | 28 | 1.94 | -5.25 | -3.31 | BGBC | 4.20 | 80 |
| CDT-2FBT-6 | 33 | 1.96 | -5.30 | -3.34 | BGBC | 5.70 | 80 |
| CDT-FBT-RA | 23.1 | 1.89 | -5.00 | -3.11 | BGBC | 17.82 | 81 |
| CDT-FBT-RR | 23.5 | 1.83 | -4.98 | -3.15 | BGBC | 9.09 | 81 |
| PCDT-FDT | 14 | 1.87 | -5.24 | -3.37 | BGBC | 9.05 | 82 |
| PCDT-FDSe | 28 | 1.57 | -5.13 | -3.56 | BGBC | 20.3 | 83 |
| IID-DT | 87.9 | 2.00 | -5.70 | -3.70 | BGTC | 0.79 | 85 |
| IID-DT-$C_3$ | 39.2 | 1.58 | -5.52 | -3.74 | BGTC | 3.62 | 86 |
| PIID-DT-Si | 138 | 1.61 | -5.20 | -3.58 | BGTC | 2.48 | 87 |
| PIID-DT-$C_6$Si | 305 | 1.60 | -5.20 | -3.60 | BGTC | 3.22 | 88 |
| PIID-DT-$C_8$Si | 434 | 1.60 | -5.20 | -3.60 | BGTC | 8.06 | 88 |
| PIID-DT | 35.6 | 1.61 | — | — | BGBC | 5.10 | 89 |
| PIID-DF-CmSi | 48.3 | 1.60 | -5.25 | -3.65 | BGTC | 4.80 | 90 |

### 3.1.3 CDT-based polymers

The cyclopentadithiophene (CDT)-based materials consisting of fused ring thiophene derivatives could lower the reorganization energy, known as a main factor that strongly affects the rate of intermolecular hopping and hence the charge carrier mobility in organic semiconductors [74]. The cyclopentadithiophene-benzothiadiazole copolymer (CDT-BTZ-$C_{16}$) thin-film devices revealed a hole mobility of 0.17 $cm^2 V^{-1} s^{-1}$, which could be further improved to a maximum of 3.3 $cm^2 V^{-1} s^{-1}$ by increasing the *M*n from 10 to 35 kDa [75]. Afterwards, CDT-BTZ-$C_{16}$ copolymer nanowires, prepared through the solvent vapor enhanced drop-casting method (SVE DC), achieved the highest hole mobility of 5.5 $cm^2 V^{-1} s^{-1}$ as a consequence of the strong intermolecular coupling between closely packed molecules and a lower density of structural defects [76]. Besides, the alkyl chain with the increased length allowed more kinetically balanced aggregation in solution, and enhanced the reorganization during compression. The maximum hole mobility of 11.4 $cm^2 V^{-1} s^{-1}$ was achieved via using CDT-BTZ-$C_{20}$ copolymer [77]. The random packing of polymer chains and the disordered arrangement of the polymer matrix strongly affected the charge mobility. To address these issues, the capillary action was applied to mediate the unidirectional alignment of polymer chains on nano-groove substrate, and the as-fabricated films based on polymers CDT-BTZ-$C_{16}$ and poly[4-(4,4-dihexadecyl-4H-cyclopenta[1,2-b:5,4-b′]-dithiophen-2-yl)-alt-[1,2,5]thiadiazolo-[3,4-c]pyridine] (CDT-PT) displayed excellent saturated hole mobilities (22.2 and 25.4 $cm^2 V^{-1} s^{-1}$ with a channel length of 80 $\mu$m, respectively) [78]. As the channel length gradually increased, the hole mobility of the polymer CDT-PT reached a maximum of 47 $cm^2 V^{-1} s^{-1}$, because the insertion of nitrogen atoms in the BT unit induced non-bonded interaction to form a quasi-linear skeleton structure and the capillaries promoted the unidirectional self-assembly of polymer chains on the nanogrooved substrate. Lately, an unprecedented ultra-high mobility ($\mu_h$ = 92.64 $cm^2 V^{-1} s^{-1}$) and air stability were found in CDT-PT nanowires via a liquid-bridge-mediated nanotransfer molding (LB-nTM) way (Fig. 5) [79]. Since the parallel alignment of CDT-PT backbone chains was along the long axis of nanowires, a minimal number of charge traps and good contact interfaces were supposed to be the main reasons for the ultrahigh mobility of single crystal CDT-PT.

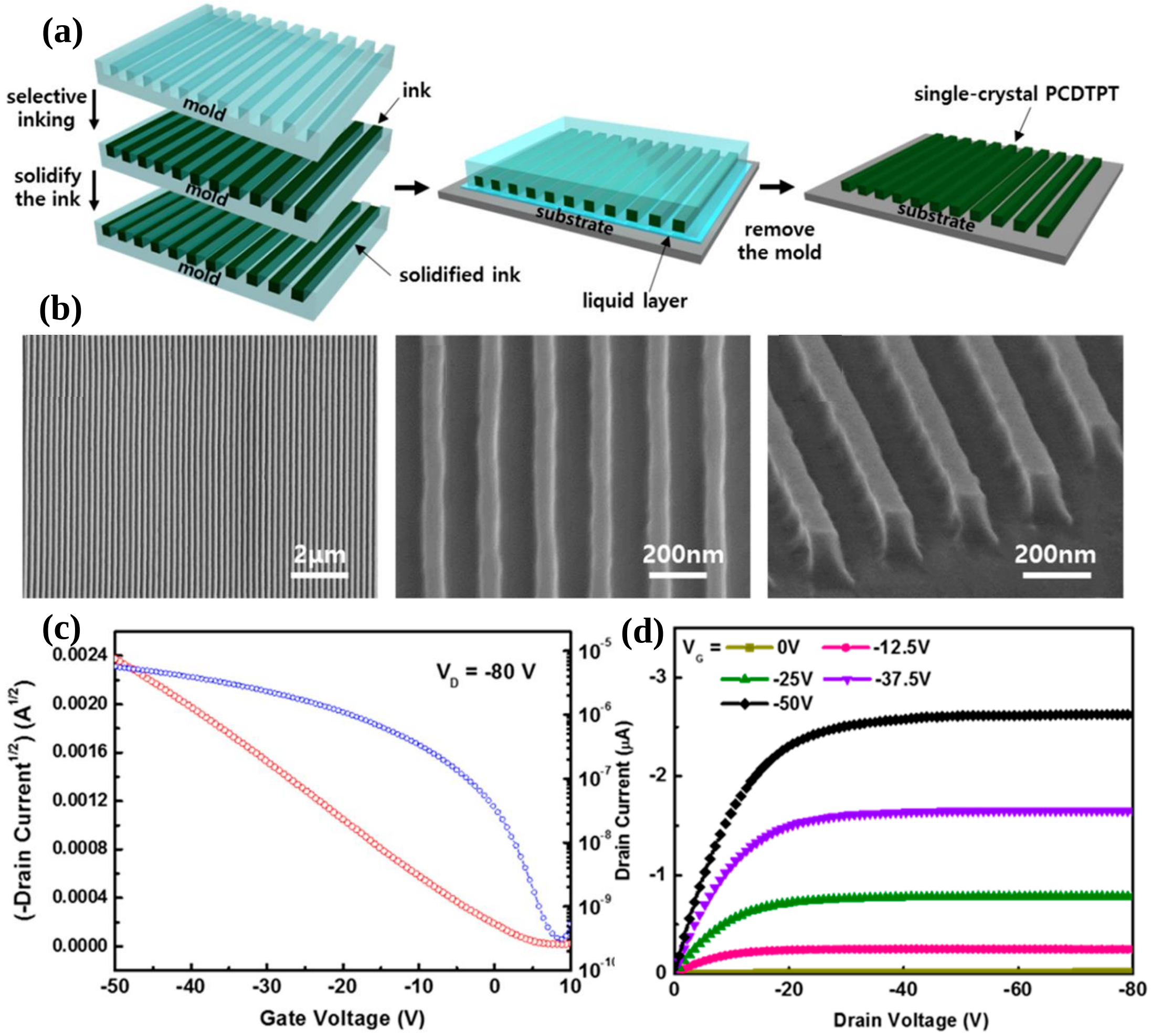

**Fig. 5** (a) Transfer and (b) output characteristics of the single-crystal CDT-PT nanowire FET. (c) Schematic illustration of the procedure used to fabricate single-crystal CDT-PT nanowires by LB-nTM. (d) SEM image of the single crystal CDT-PT nanowires (white) on Si substrates (black) and a partial enlarged view. Reproduced from Ref. [79].

In order to improve the planarity of the CDF-2FBT backbone, three new cyclopentadithiophene-difluorophenylene copolymers (CDF-2FBT-3, CDF-2FBT-5 and CDF-2FBT-6) were synthesized, where the different arrangement of fluorines on the phenylene structural units generated variable non-bonding F···H interactions, which impacted the secondary structure of backbones [80]. For linear polymers CDF-2FBT-5 and CDF-2FBT-6, due to their highly anisotropic alignment along the nanogrooved (NG) direction, significantly improved charge carrier mobilities of 4.2 and 5.7 cm$^2$ V$^{-1}$ s$^{-1}$ on NG substrates were determined, comparing to the zig-zag CDF-2FBT-3 polymer ($\mu_h$ = 2.8 cm$^2$ V$^{-1}$ s$^{-1}$). Recently, regiorandom (RA) and regioregular version (RR) cyclopenta[1,2-b:5,4-b]dithiophene fluorobenzo[c][1,2,5]thiadiazole polymer (CDT-FBT) were investigated to possess similar optical, electrochemical and morphological properties, but different FET characteristics [81]. By using the sandwich-casting technique on a NG substrate, the hole mobilities of the polymers

CDT-FBT-RA and CDT-FBT-RR were significantly improved to 17.82 cm$^2$ V$^{-1}$ s$^{-1}$ and 9.09 cm$^2$ V$^{-1}$ s$^{-1}$, respectively. In addition, a conjugated polymer poly(4-(4,4-bis(2-ethylhexyl)-4*H*-silolo[3,2-*b*:4,5-*b*′]dithiophen-2-yl)-7-(4,4-bis(2-ethylhexyl)-6-(thiophen-2-yl)-4*H*-silolo[3,2-*b*:4,5-*b*′]dithiophen-2-yl)-5,6-difluorobenzo[*c*][1,2,5]thiadiazole) (PCDT-FDT), constructed from a derivative of CDT and fluorinated BT unit, showed three different hole mobilities of 2.6, 2.8, and 9.0 cm$^2$ V$^{-1}$ s$^{-1}$, which were extracted from top gate/bottom contact (TG/BC) OTFTs with common poly(methylmethacrylate) (PMMA), high-*k* poly(vinylidenefluoride–trifluoroethylene–chlorotrifluoroethylene) P(VDF–TrFE–CTFE) and poly-(vinylidenefluoride–trifluoroethylene) P(VDF–TrFE) as dielectric insulators, respectively [82]. Through replacing S atom in the FDT unit with Se atom, polymer PCDT-FDSe film had the well-packed highly-crystalline state (unique edge-on) and displayed the hole mobility up to 20.3 cm$^2$ V$^{-1}$ s$^{-1}$ by fabricating TG/BC organic thin-film transistor with solid electrolyte gate insulator (SEGI) of P(VDF-TrFE)/poly(vinylidene fluoride-hexafluoropropylene) [83]. In contrast to PCDT-FDT films, crystalline domains of PCDT-FDSe films are predominantly edge-on orientation, which is beneficial for carrier transport in TFT device geometries.

### 3.1.4 IID-based polymers

The isoindigo (IID) unit has several advantageous characteristics such as extended conjugation length, large dipole moment and high planarity [84]. The first reported IID copolymer (DT as the donor unit) based field-effect transistor showed a hole mobility as high as 0.79 cm$^2$ V$^{-1}$ s$^{-1}$and good stability under high humidity of environmental conditions [85]. Furthermore, the branch point length of the alkyl side chains affected the FET performance of the IID-based conjugated polymer. With increased branch point length, IID-DT-$C_3$ thin-film adopted a long-range orderly edge-on orientation, a shorter $\pi$–$\pi$ stacking distance (3.57 Å), and a higher mobility of 3.62 cm$^2$ V$^{-1}$ s$^{-1}$ than IID-DT (1.06 cm$^2$ V$^{-1}$ s$^{-1}$) and IID-DT-$C_2$ (0.40 cm$^2$ V$^{-1}$ s$^{-1}$) [86]. The introduction of siloxane-terminated solubilizing groups allowed the conjugated polymers to have sufficient solubility for solution-processable treatment. The interchain $\pi$-$\pi$ stacking distance in PIID-DT-Si polymer thin-film was about 3.58 Å, smaller than 3.76 Å of referenced (PIID-DT-Ref) thin-film, leading to the increased hole mobility of 2.48 cm$^2$ V$^{-1}$ s$^{-1}$ [87]. It is well known that conjugated polymers with branched alkyl groups containing differed linear carbon spacers affect their field-effect performances. However, long and branched alkyl chains are not easily synthesized and the yield is generally low. Thus, carbosilane side chain is introduced to solve this issue. The PIID-DT-$C_6$Si and PIID-DT-$C_8$Si polymers with various carbosilane side chains had the ordered solid-state molecular stacking, excellent charge transport characteristics (PIID-DT-$C_6$Si: $\mu_h$ = 3.22 cm$^2$ V$^{-1}$ s$^{-1}$, PIID-DT-$C_8$Si: $\mu_h$ = 8.06 cm$^2$ V$^{-1}$ s$^{-1}$) in the spin-coating thin-films, and the branch point in the side chain far away from the conjugated backbone chains induced denser packing modes between polymer chains [88]. Nevertheless, another main reason to introduce such a huge carbosilane side group is to improve the mechanical property of the semiconducting polymer for wearable/stretchable electronic applications, because the long and floppy side chain reduced the volume fraction of the brittle conjugated backbone and weakened the elastic modulus. The two polymer thin films, prepared via the transfer method, presented stable and reliable charge transport abilities under an applied tensile strain up to 100% with a charge transport direction parallel or perpendicular to the strain direction (Fig. 6). Besides, symmetry breaking in side chains is an effective method to alter the molecular orientation, enabling the isoindigo-alt-bithiophene (PIID-DT) polymer with asymmetric side chains to adopt a distinct bimodal orientation and close interchain packing in the thin-film with a high hole mobility of 5.1 $cm^2\ V^{-1}\ s^{-1}$. At the meanwhile, this PIID-DT polymer had excellent solubility in various organic solvents [89]. Moreover, a series of new IID-based conjugated polymers (PIID-DF-$C_m$Si, $m$ = 3–11) with alkyl siloxane-terminated side chains were prepared by changing the length of the main chain and the linear alkyl chain spacing [90]. As the length of the branching point increased from PIID-DF-$C_3$Si to PIID-DF-$C_9$Si, the polymer $\pi$-$\pi$ stacking distance (3.38-3.41 Å) became shorter, corresponding to the highest hole mobility of 4.8 $cm^2\ V^{-1}\ s^{-1}$. However, when the length of the branches continued to increase, the longer and floppy side chains would cause the slow crystallization and chaotic crystallite orientation.

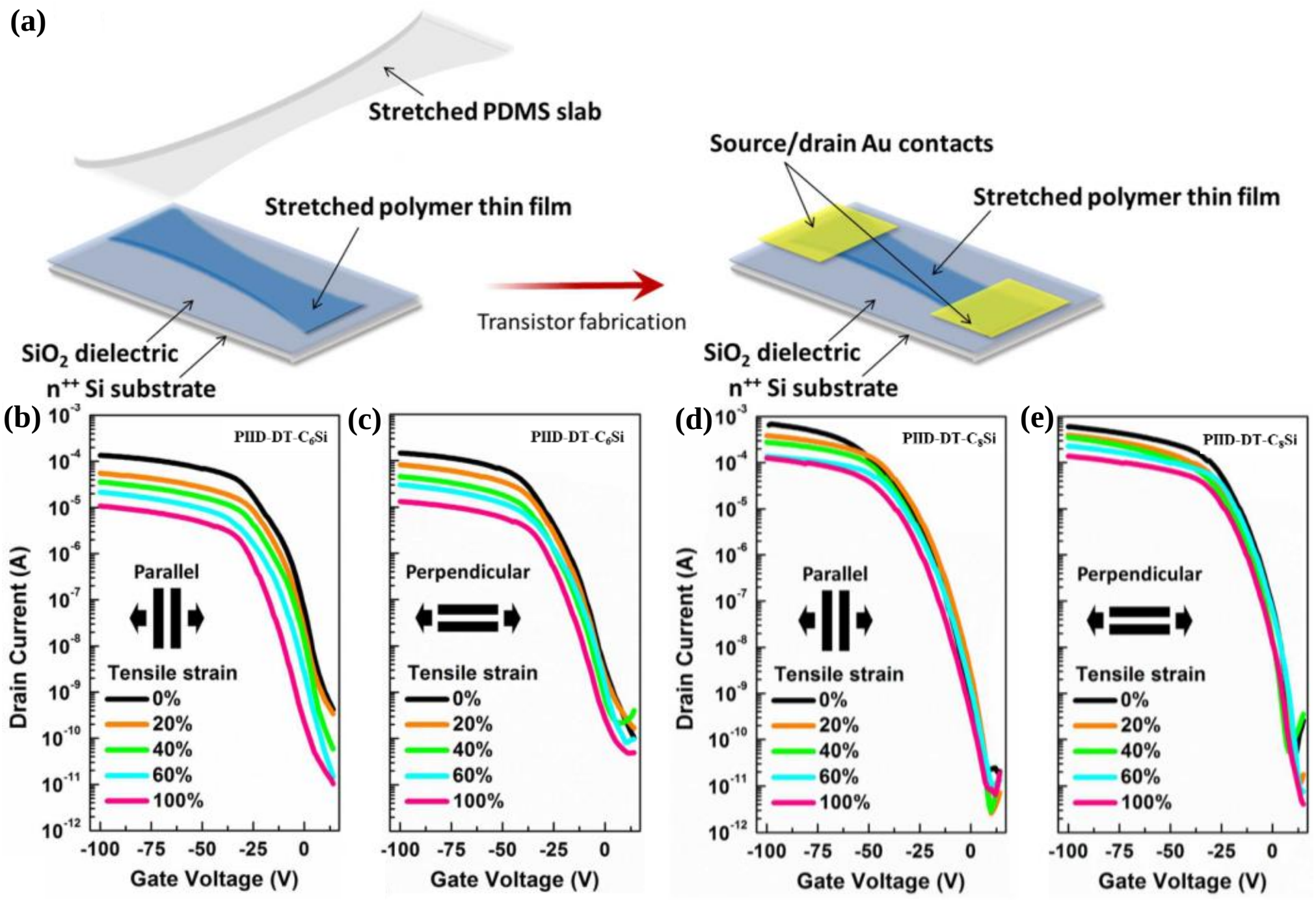

**Fig. 6** (a) Schematic illustration of fabrication of a FET device based on the studied semiconducting polymer thin films under strain. FET transfer characteristics of (b, c) IID-DT-$C_6$Si and (d, e) IID-DT-$C_8$Si-based stretched films with a charge transport direction (b, d) parallel or (c, e) perpendicular to the strain direction, respectively. Reproduced from Ref. [88].

## 3.2 N-Type Conjugated Polymers

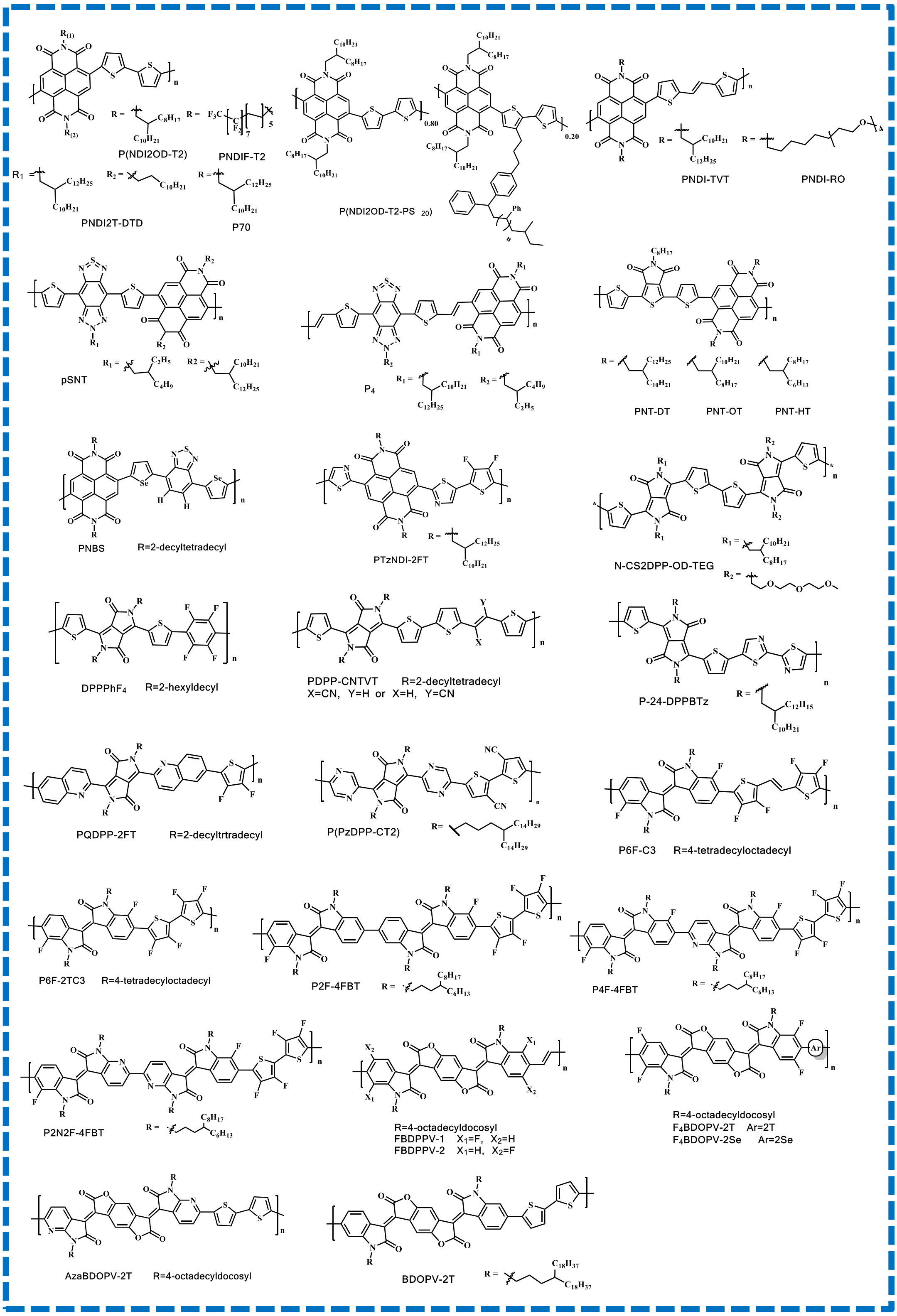

**Fig. 7** The structures of some high-performance n-type conjugated polymers

Since n-type organic thin-film transistors are the key part to construct low-power complementary metal oxide semiconductor (CMOS)-like logic circuits, significant progress in developing novel n-type conjugated polymers has been witnessed in the last two decades. However, the performance of n-type conjugated polymers lags behind that of their p-type counterparts mainly due to the high LUMO level [91]. Aiming at improving electron-transport performance, many researchers have attempted to adjust the packing fashion and/or frontier molecular orbital energy levels through molecular design strategies [92]. This part mainly summarizes the high-performance n-type conjugated polymers in recent years (Fig. 7 and Table 3).

Table 3. Molecular weight, optical and electrochemical properties, OFET device structure, and performance of NDI-, DPP-, IID-, BDOPV-based high-performance n-type conjugated polymers.

| Polymer | *M*n[kDa] | $E_g^{opt}$[eV] | HOMO[eV] | LUMO[eV] | Device structure | $\mu_h$,max [$cm^2 V^{-1} s^{-1}$] | Refs |
|---|---|---|---|---|---|---|---|
| P(NDI2OD-T2) | 52.5 | 1.45 | −5.36 | −3.91 | TGBC | 0.85 | 93 |
| P(NDI2OD-T2-$PS_{20}$) | 29.2 | 1.45 | -5.33 | -3.88 | BGTC | 0.30 | 96 |
| P70 | 17.18 | 1.85 | -5.80 | -3.95 | BGTC | 0.77 | 98 |
| PNDIF-T2 | 28 | 1.61 | -5.62 | -4.01 | BGTC | 6.50 | 99 |
| PNDI-TVT | 70 | 1.42 | -5.42 | -4.00 | TGBC | 1.80 | 100 |
| PNDI-RO | 24 | 1.32 | -5.30 | -4.01 | BGTC | 1.64 | 101 |
| PNDI2T-DTD | — | 1.41 | -5.81 | −3.91 | BGTC | 1.52 | 102 |
| pSNT | 61.3 | 1.14 | -5.45 | -3.88 | BGTC | 5.35 | 103 |
| P4 | 54.9 | 1.08 | −5.4 | −3.87 | BGTC | 7.16 | 104 |
| PNT-DT | 29.3 | 1.70 | -6.06 | -3.94 | BGTC | 0.38 | 105 |
| PNT-OD | 21.6 | 1.65 | -6.03 | -3.95 | BGTC | 0.68 | 105 |
| PNT-HD | 20.4 | 1.58 | -5.97 | -4.01 | BGTC | 1.05 | 105 |
| PNBS | 39.7 | 1.03 | -5.84 | -3.81 | TGBC | 8.50 | 106 |
| PTzNDI-2FT | 79.4 | 1.72 | −6.24 | −3.89 | TGBC | 0.57 | 107 |
| N-CS2DPP-OD-TEG | 67.3 | — | -5.38 | -3.66 | BGBC | 3.00 | 109 |
| $DPPPhF_4$ | 16.3 | 1.47 | −5.65 | −4.18 | BGTC | 2.36 | 110 |
| PDPP-CNTVT | 125.2 | 1.27 | −5.77 | −3.92 | TGBC | 7.00 | 111 |
| P-24-DPPBTz | 56 | 1.35 | -5.31 | -3.91 | TGBC | 1.87 | 112 |
| PQDPP-2FT | 67.4 | 1.80 | -5.64 | -3.84 | TGBC | 6.04 | 113 |
| P(PzDPP-CT2) | 28.5 | 1.86 | -5.89 | -4.03 | TGBC | 0.79 | 114 |
| PIIG-BT | 25 | — | –5.70 | –3.54 | BGTC | 0.22 | 117 |
| PAIIDBT | 14 | — | -5.50 | -3.40 | TGBC | 1.00 | 118 |
| P6F-C3 | 52.9 | 1.63 | -5.95 | -3.80 | TGBC | 4.97 | 119 |
| P6F-2TC3 | 88 | 1.63 | -6.09 | -3.92 | TGBC | 1.35 | 119 |
| P2F-4FBT | 48.9 | — | -5.95 | -3.89 | TGBC | 0.93 | 120 |
| P4F-4FBT | 17 | — | -6.06 | -3.99 | TGBC | 0.71 | 120 |
| P2N2F-4FBT | 20.3 | — | -6.04 | -4.01 | TGBC | 1.24 | 120 |
| FBDPPV-1 | 66.3 | 1.46 | −6.19 | −4.26 | TGBC | 1.70 | 122 |
| FBDPPV-2 | 53.8 | 1.39 | −6.22 | -4.30 | TGBC | 0.81 | 122 |
| $F_4$BDOPV-2T | 38 | 1.31 | –5.96 | –4.32 | TGBC | 14.9 | 123 |
| $F_4$BDOPV-2Se | 23.8 | 1.29 | –5.91 | –4.34 | TGBC | 6.14 | 123 |
| AzaBDOPV-2T | 51.6 | — | -5.80 | -4.37 | TGBC | 3.22 | 124 |
| BDOPV-2T | 77.2 | 1.32 | -5.59 | -4.27 | BGTC | 3.20 | 125 |

### 3.2.1 NDI-based polymers

The NDI core has been identified as the commonly building block in high-performance n-type polymers due to its relatively low LUMO energy level (−3.6 eV). A typical NDI-based polymer (P(NDI2OD-T2)) exhibited a high unipolar electron mobility ($\mu_e$) up to 0.85 cm$^2$ V$^{-1}$ s$^{-1}$ as well as remarkable ambient stability with TGBC structure [93]. Lately, a method, named as kinetically controlled crystallization (KCC) from nano to microscale, was reported to improve the electron transport performance of P(NDI2OD-T2)-based thin-film devices (3.99 cm$^2$ V$^{-1}$ s$^{-1}$), which was closely related to large domains, distinct fibrillar structure, and favorable alignment optimized for efficient charge transport [94]. Moreover, because of the high reduction potential, benzyl viologen (BV) was used as a molecular N-dopant to induce direct intermolecular charge transfer [95]. When the BV dopant concentration increased from 0.5 wt% to 2.0 wt%, the P(NDI2OD-T2)-based FETs could reach a maximum electron mobility of 1.8 cm$^2$ V$^{-1}$ s$^{-1}$, attributing to the increased charge carrier density in the semiconductor channel through a direct redox reaction (charge transfer) between BV and P(NDI2OD-T2). By tuning the monomer ratio in the random copolymer, low molecular-weight polystyrene (PS) oligomer side chains was incorporated into NDI-based conjugated polymers to improve the performance of the NDI-based polymer [96]. With 20 mol% of PS side chain, the highest mobility of P(NDI2OD-T2-PS$_{20}$) could reach 0.3 cm$^2$ V$^{-1}$ s$^{-1}$, proving that the PS side chains could act as a molecular encapsulation layer around the conjugated polymer backbone, leading to its less susceptibility to electron traps. Besides, as a nonvolatile n-type molecular dopant, (12a,18a)-5,6,12,12a,13,18,18a,19-octahydro5,6-dimethyl-13,18[1′,2′]-benzenobisbenzimidazo[1,2-b:2′,1′ d]benzo[i][2.5]benzodiazocine potassium triflate adduct (DMBI-BDZC) has recently been doped into P(NDI2OD-T2)-based polymer thin film to enhance the charge transfer of organic semiconductor, the improvement of electron mobility (up to 0.7 cm$^2$ V$^{-1}$ s$^{-1}$) and a systematic threshold voltage reduction (35 V vs 10 V) were found due to the trap of filling states [97].

A naphthalenediimide-dithiophene conjugated polymer (P70), synthesized by continuous flow synthesis, displayed an electron mobility of 0.77 cm$^2$ V$^{-1}$ s$^{-1}$ on account of higher molecular-weight (17.28/35.27 vs. 4.86/13.21 kg mol$^{-1}$), lower polydispersity indexes (2.0 vs. 2.7), and a more ordered film crystalline compared to flask synthesis [98]. Moreover, the introduction of fluoroalkyl chains in organic semiconductors has been developed because of the fluorophobic interactions. A −$(CH_2)_{10}(CF_2)_7CF_3$ semifluorinated alkyl side chain was employed into NDI-based polymer to prepare poly[(*E*)-2,7-bis(11,11,12,12,13,13,14,14,15,15,16,16,17,17,18,18,18-heptadecafluorooctadecyl)-4-methyl-9-(5-(2-(5-methylthiophen-2-yl)vinyl)thiophen-2-yl)benzo[lmn][3,8]phenanthroline-1,3,6,8(2*H*,7*H*)-tetraone] (PNDIF-T2) [99]. With the addition of a higher boiling point solvent 1-chloronaphtalene (CN), the bar-coated PNDIF-T2 films with superior crystalline order have shown very high electron mobilities up to 6.5 cm$^2$ V$^{-1}$ s$^{-1}$, mainly owing to the improved chain packing. In addition, efforts have also been made to alter the donor unit from T2 to TVT to provide better charge-transporting properties. For instance, poly[(*E*)-2,7-bis(2-decyltetradecyl)-4-methyl-9-(5-(2-(5methylthiophen-2-yl)vinyl)thiophen-2-

yl)benzo[lmn][3,8]phenanthroline-1,3,6,8(2*H*,7*H*)-tetraone] (PNDI-TVT) based top-gate, bottom-contact device with $Cs_2CO_3$-treated source/drain Au electrodes displayed a $\mu_e$ value of 1.8 $cm^2 V^{-1} s^{-1}$ and good air stability because of the crystallinity enhancement by the extended *π*-*π* backbone and long alkyl groups [100]. Upon incorporating the combination of hydrophobic alkyl chain (–R–) and hydrophilic oligo(ethylene glycol) chain (–O–), NDI and TVT units could be alternately coupled together to obtain a new copolymer poly[(*E*)-4-methyl-9-(5-(2-(5-methylthiophen-2-yl)vinyl)thiophen-2-yl)-2,7-di(2,5,8,11-tetraoxanonadecan-19-yl)benzo[*lmn*][3,8]phenanthroline-1,3,6,8(2*H*,7*H*)-tetraone] (PNDI-RO) with superior polymer-chain rigidity [101]. Interestingly, the electron mobility of PNDI-RO-based thin-film device was improved from 0.51 to 1.6 $cm^2 V^{-1} s^{-1}$ by using a CN additive, which could be attributed to the good crystallinity, resulting from the reduced evaporation rate of the mixed solvent. Linear side chains in the polymer backbone was found to play a great role in the conjugated backbone planarity for good thin-film crystallinity. A new soluble NDI-T2 copolymer poly[2-(2-decyltetradecyl)-7-dodecylbenzo-2,6-dibromo-1,4,5,8-naphthalenediimide-bithiophene] (PNDI2T-DTD) with a long branching alkyl chain and a linear alkyl chain at the *N*, *N'*-position of each NDI unit demonstrated electron mobility up to 1.52 $cm^2 V^{-1} s^{-1}$ with the promotion of interchain interdigitation and the planarity of the conjugated backbone [102].

Recently, a "dual-acceptor" strategy to combine two acceptor building blocks into one polymer repeat unit was applied to design the novel n-type conjugated polymers. For instance, a new D-$A_1$-D-$A_2$ type polymer pSNT, composed of NDI and 4,7-bis(5-(trimethylstannyl)thiophen-2-yl)-[1,2,5]thiadiazolo[3,4-*f*]benzotriazole (SNT), displayed highest $\mu_e$ of 4.87 $cm^2 V^{-1} s^{-1}$ and weak p-type character ($\mu_h$ = 1.1 × $10^{-2}$ $cm^2 V^{-1} s^{-1}$), while through the surface modification of [3-(*N*,*N*-dimethylamino)propyl]trimethoxysilane (NTMS), unipolar n-type characteristics (5.35 $cm^2 V^{-1} s^{-1}$) was realized, in which the amine groups could trap the holes and promote electron accumulation [103]. However, these "dual-acceptor" type polymers easily suffered from the problems of short coherent conjugation length and poor intrachain delocalization of the HOMO/LUMO orbitals due to the twisted polymer backbone. Subsequently, Wang et al. successfully synthesized a new highly planar backbone polymer P4 through inserting vinylene bridges in pSNT backbone, which can form intramolecular hydrogen bonds to tune the coplanar backbone conformation [104]. By using NTMS as the SAM layer, the P4-based thin-film transistors showed a high $\mu_e$ of 7.16 $cm^2 V^{-1} s^{-1}$. To further optimize the D-$A_1$-D-$A_2$ polymers electron-transporting properties, the side-chain engineering strategy was developed. When the side-chain length became shorter from 2-decyltetradecyl (DT) to 2-octadecyldodecyl (OD) and further to 2-hexyldecyl (HD), the D−$A_1$−D−$A_2$ type of the as-prepared poly(naphthalenediimide-thieno[3,4-*c*]pyrrole-4,6(5*H*)-dione) (PNT) derivatives (*i.e.* PNT-DT, PNT-OD and PNT-HD) were found to show the increased electron mobilities from 0.38, 0.68 to 1.05 $cm^2 V^{-1} s^{-1}$ owing to the advancement of intermolecular interactions between the backbones and deeper energy levels [105].

Due to the large p-orbital size of selenium atom in selenophene heterocycle substitution, the naphthalenediimide-selenophene copolymer (PNBS) thin-films, prepared through simple spin-coating technique, had smooth surfaces and close $\pi$–$\pi$ stacking distance. After pentafluorobenzenethiol (PFBT) modification, the work function of Au electrodes was well tuned from 5.13 eV (bare Au) to 4.77 eV (PFBT-Au), engendering high maximum electron mobility of the OTFT devices with a thick poly-methyl methacrylate (PMMA, 1350 nm) dielectric layer up to 8.5 $cm^2\ V^{-1}\ s^{-1}$ (Fig. 8) [106].

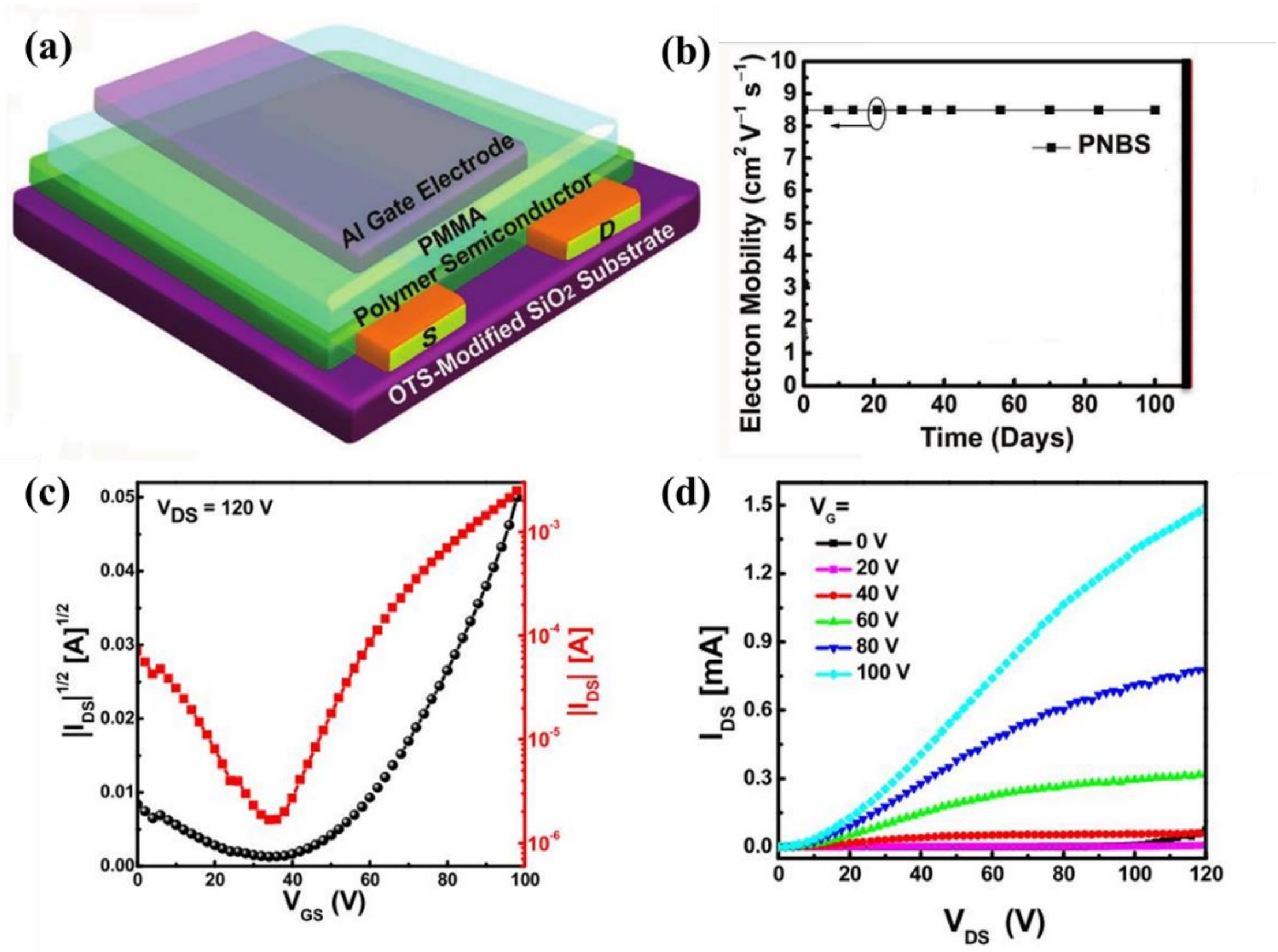

**Fig. 8** (a) TGBC device structure for PNBS-based polymer FETs. (b) Air-stability of PNBS-based FETs in ambient conditions. (c) Transfer and (d) output characteristics of PNBS-based TGBC FET devices fabricated on glass substrates. Reproduced from Ref. [106].

A thiazole-flanked NDI unit (TzNDI), designed to simultaneously lower HOMO energy level and improve the flatness of polymer backbone, was used to construct a conjugated polymer poly(TzNDI-3,4-difluorothiophene) (PTzNDI-2FT) with high molecular weight and excellent thermal stability, affording PTzNDI-2FT-based OFETs a high $\mu_e$ of 0.57 $cm^2\ V^{-1}\ s^{-1}$, higher than that of P(NDI2OD-T2) (0.41 $cm^2\ V^{-1}\ s^{-1}$) under the same conditions [107].

### 3.2.2 DPP-based polymers

At present, many studies have focused on further lowering the LUMO energy level of DPP-based polymers by introducing strong electron-withdrawing groups to achieve the conversion from p-type to n-type transport [108]. The copolymer poly(3,6-bis(5-(4,4,5,5-tetramethyl-1,3,2dioxaborolan-2-yl)thiophen-2-yl)-*N*,*N*-bis(2-octyldodecyl)-1,4dioxopyrrolo[3,4-*c*]pyrrole-3,6-bis(5-bromothiophen-2-yl)-*N*,*N*-bis(2-(2-(2-methoxyethoxy)ethoxy)ethyl)-1,4dioxopyrrolo[3,4-*c*]pyrrole) (N-CS2DPP-OD-TEG), produced by coupling of DPP with DPP in an alternating fashion, exhibited the field-effect electron mobility values of 3 $cm^2\ V^{-1}\ s^{-1}$, which could be imputed to the strong intermolecular donor-acceptor interaction between DPP units [109]. Besides, as the number of fluorine substitutions increased, the LUMO energy level of fluorine-substituted DPP-based donor-acceptor conjugated polymer poly(3,6-bis(5-bromo-2-thienyl)-2,5-dihydro-2,5-di(2-hexyldecyl)-pyrrolo[3,4c]pyrrolo-1,4-dione-2,3,5,6-tetrafluoro-1,4-phenylene(bis(triemethyltin))) ($DPPPhF_4$) reduced as low as -4.18 eV, and the face-on orientation of $DPPPhF_4$ thin-film became favorable, leading to a p-to-n type switch of the charge transport behavior (2.36 $cm^2\ V^{-1}\ s^{-1}$) [110]. However, the torsion angle ($\varphi$) between the thiophene and the phenylene ring in $DPPPhF_4$ was somewhat large, although the planarity of the conjugated backbone was obviously improved due to the fluorination of DPP-based conjugated backbones. The attached position of electron-withdrawing groups (EWGs) should be carefully optimized to minimize the steric hindrance effect during the substitution. By the incorporation of the nitrile group as an electron-withdrawing unit in the vinyl linkage during the substitution, a DPP-based polymer poly[2,5-bis(2-octyldodecyl)pyrrolo[3,4-*c*] pyrrole-1,4(2*H*, 5*H*) dione-(*E*)-[2,2'-bithiophen]-5-yl)-3-(thiophen-2-yl)acrylonitrile] (PDPP-CNTVT) with the LUMO energy level of -3.92 eV exhibited a $\mu_e$ up to 7.00 $cm^2\ V^{-1}\ s^{-1}$ in thin-film devices [111]. Interestingly, the device characteristics of PDPP-CNTVT OFETs continually increased with the increased PDPP-CNTVT film thickness (Fig. 9a), which could be related to the large number of electron traps in the entire bulk of PDPP-CNTVT film (Fig. 9b). Besides, the diketopyrrolopyrrole-thiophene-bithiazole polymer (P-24-DPPBTz) containing branched alkyl chains of 24-alkyl with C1 spacer was synthesized, and showed n-type dominant electrical properties with electron mobility of 1.87 $cm^2\ V^{-1}\ s^{-1}$ on account of the strong electron deficiency and trans-planar conformation of bithiazole group [112].

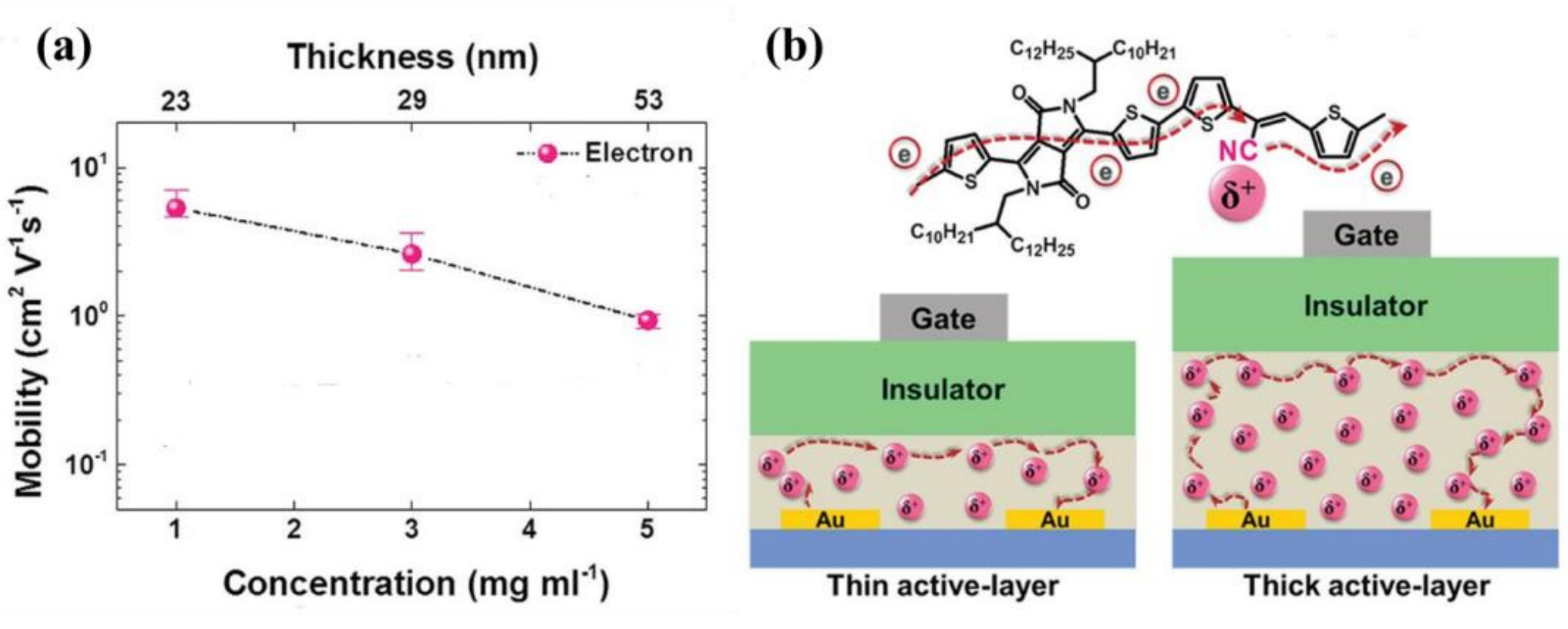

**Fig. 9** (a) Mobility varies with film thickness and concentration. (b) Schematic illustration of electron charge transport and trapping incidents in PDPP-CNTVT layers with different active-layer thicknesses. Reproduced from Ref. [112].

Copolymerizing $\pi$-extended quinoline-flanked DPP and perfluorinated thiophene units could produce PQDPP-2FT, which exhibited an electron mobility of its thin-film devices up to 6.04 $cm^2\ V^{-1}\ s^{-1}$ [113]. This high mobility resulted from a high planarity of conjugated backbone and close $\pi$-$\pi$ stacking (≈ 3.45Å), enabled by the non-bonded interaction between the fluorine atom of perfluorinated thiophene and the hydrogen atom of quinolone. Recently, with the deepest LUMO level in all the reported DPP derivatives, a new donor-acceptor copolymer derivative-pyrazine flanking DPP-3,3′dicyano-2,2′-bithiophene (P(PzDPP-CT2)) was designed and synthesized [114]. After mixed with n-dopant (4-(1,3-dimethyl-2,3dihydro-1*H*-benzoimidazol-2-yl)phenyl)dimethylamine (N-DMBI), the P(PzDPP-CT2) film showed good thermoelectric properties of a high n-type conductivity of 8.4 S $cm^{-1}$ and a power factor up to 57.3 W $m^{-1}\ K^{-2}$, as well as a high electron mobility of 0.79 $cm^2\ V^{-1}\ s^{-1}$.

### 3.2.3 IID-based polymers

Because the LUMO energy level of the IID unit is relatively low (-3.5 eV), it has been widely considered as one of the most commonly-used electron-deficient core units in high-performance conjugated polymers [115]. Many studies have proven that further reduction of the LUMO energy level of IID-based polymers through chemical modification can improve electron transport and device performance [116].

An isoindigo-benzothiadiazole acceptor-acceptor copolymer (PIIG-BT) was synthesized via Suzuki coupling, demonstrating an electron mobility up to 0.22 $cm^2\ V^{-1}\ s^{-1}$, mainly due to the very low-lying HOMO level (-5.70 eV) of the A–A type IIG-based copolymer [117]. By incorporating a highly electron-deficient azaisoindigo core, the A-A type azaisoindigo-benzothiadiazole copolymer (PAIIDBT) with both lower LUMO and HOMO energy levels at -4.1 eV and -5.8 eV exhibited an electron mobility as high as 1 $cm^2\ V^{-1}\ s^{-1}$ [118]. DFT simulations suggested that the non-bonding interaction, formed through nitrogen replacement at the 7-position of the azaisoindigo unit, significantly reduced the steric hindrance between structural units to improve the planarity of the skeleton, which might explain the high carrier mobility of PAIIDBT thin-films.

A "multifluorination" method, that fluorine atoms incorporated into both D and A units to reduce the frontier orbital energy level and enhanced the coplanarity of conjugated backbones, has been applied [119]. The as-prepared IID-based polymers isoindigo-(*E*)-1,2-bis(3,4-difluorothien-2-yl)ethene (P6F-C3) and isoindigo-3,3′,4,4′-tetrafluoro-2,2′-bithiophene (P6F-2TC3), bearing close $\pi$-$\pi$ stacking distance (3.43 and 3.47 Å) of two copolymers, illustrated the electron mobility of 4.97 $cm^2\ V^{-1}\ s^{-1}$ and 1.35 $cm^2\ V^{-1}\ s^{-1}$, respectively. Recently, a series of n-type conjugated polymers (P2F-4FBT, P4F-4FBT, P2N2F-4FBT) containing different numbers of electron withdrawing fluorine (F) or nitrogen (N) atoms were synthesized by copolymerizing 6,6‴-dibromo-7,7‴-difluoro-N,N′,N″,N‴-tetrakis(4hexyldodecyl)-6′,6″-bisisoindigo, 6,6‴-dibromo7,7′,7″,7‴-tetrafluoro-N,N′,N″,N‴-tetrakis(4-hexyldodecyl)6′,6″-bisisoindigo, and 6,6‴-dibromo-7,7‴-difluoroN,N′,N″,N‴-tetrakis(4-hexyldodecyl)-6′,6″-bisazaisoindigo with 3,3′,4,4′-tetrafluoro2,2′-bithiophene to finely tune HOMO and LUMO energy levels, and displayed electron mobilities of 0.93, 0.71, and 1.24 $cm^2 V^{-1} s^{-1}$, respectively [120].

### 3.2.4 BDOPV-based polymers

BDOPV unit can also be employed as acceptor units in n-type conjugated polymers, owing to its strong electron-deficient property [121]. Through fluorine atom substitution, two high-performance n-type benzodifurandione-based poly(p-phenylene vinylene) (BDPPV) polymers FBDPPV-1 and FBDPPV-2 with the LUMO energy levels as low as −4.30 eV were obtained ($\mu_e$ of FBDPPV-1 = 1.70 $cm^2 V^{-1} s^{-1}$; $\mu_e$ of FBDPPV-2 = 0.81 $cm^2 V^{-1} s^{-1}$) [122]. Fluorination endowed lower LUMO levels, more ordered thin-film packing, smaller $\pi-\pi$ stacking distance, stronger interchain interaction, and locked conformation of polymer backbones. In another study, the fluorinated BDOPV derivatives copolymerized with bithiophene or biselenophene to form $F_4$BDOPV-2T and $F_4$BDOPV-2Se, both of which showed the improved $\mu_e$ values of 14.9 $cm^2 V^{-1} s^{-1}$ and 6.14 $cm^2 V^{-1} s^{-1}$, respectively, with low-lying LUMO energy levels (-4.32 eV and -4.34eV, respectively) and high backbone planarity through nonbonding interactions [123]. In addition, with $sp^2$-nitrogen atoms embedded in an isatin unit, a D-A conjugated polymer diaza-benzodifurandione-based oligo(p-phenylene vinylene)-2,2'-bithiophene (AzaBDOPV-2T) with a LUMO energy level as low as -4.37 eV and conformation-locked planar backbone was developed to display an electron mobility of 3.22 $cm^2 V^{-1} s^{-1}$ [124].

Recently, a “solution-state supramolecular structure control” strategy by using co-solvent system to tune the solution-state structure from 1D rod-like in a good solvent (1,2-dichlorobenzene) to 2D layered in a poor solvent (toluene) was employed to alter the electron mobility of benzodifurandione-based oligo(p-phenylene vinylene)-2,2'-bithiophene polymer (BDOPV-2T). Through adding 20% toluene in 1,2-dichlorobenzene, the as-prepared spin-coated thin-films with both high crystallinity and good inter-domain connectivity exhibited the electron mobility of 3.2 $cm^2 V^{-1} s^{-1}$, while the BDOPV-2T thin film from pure 1,2-dichlorobenzene solution showed electron mobility of 1.8 $cm^2 V^{-1} s^{-1}$, resulting from the further polymer aggregation induced by poor solvents [125].

## 3.3 Ambipolar Conjugated Polymers

Ambipolar conjugated polymers, which exhibit both holes and electrons transport features, have acted as a pivotal part in low-cost fabrication of complementary-like inverters. In this section, we will review the well-balanced ambipolar polymers with high charge carrier mobilities and illustrate design strategies for polarity control of the conjugated polymers (Fig.

10 and Table 4).

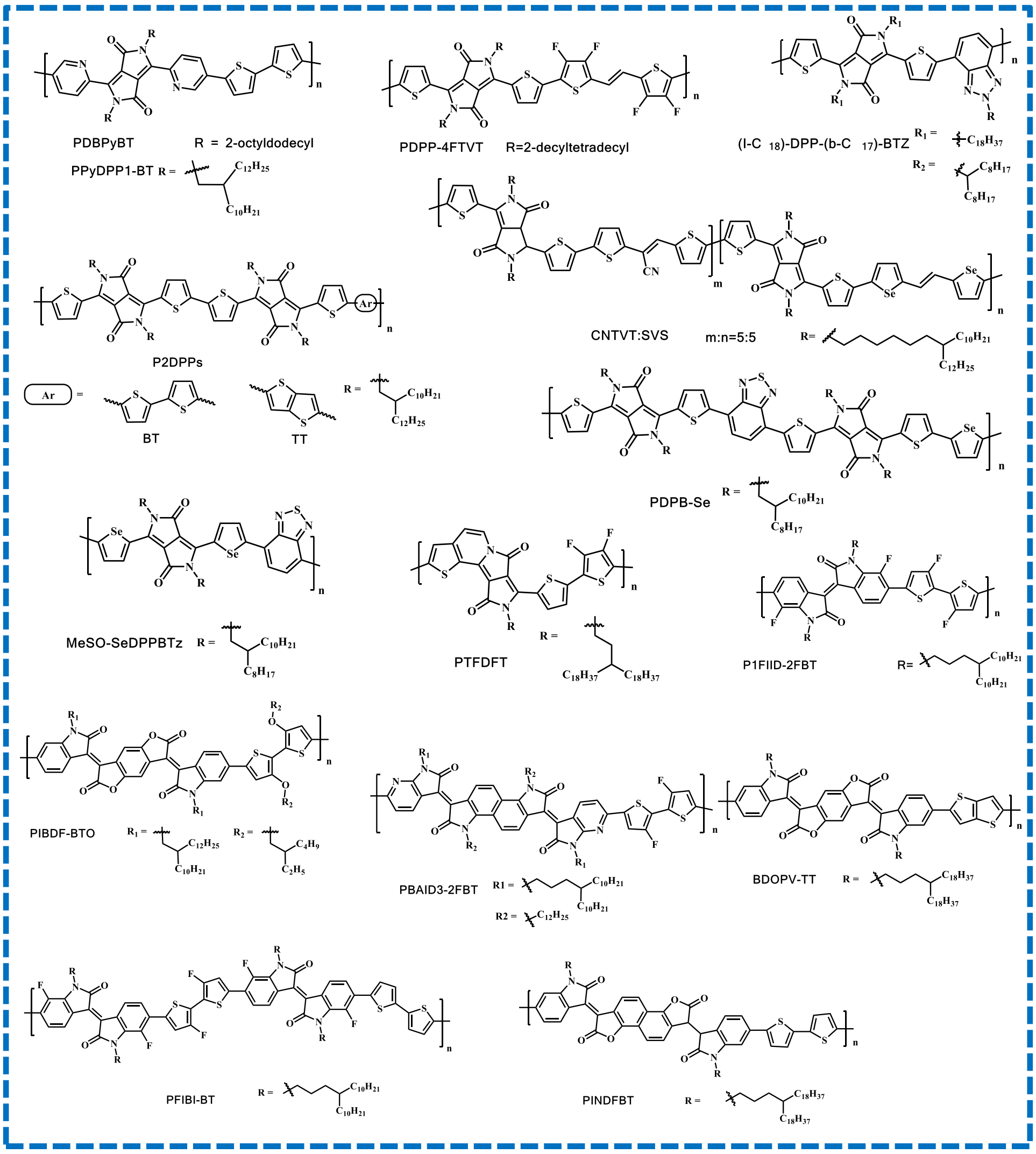

**Fig. 10** High-performance ambipolar conjugated polymers

Table 4. Molecular weight, optical and electrochemical properties, OFET device structure, and performance of high-performance ambipolar conjugated polymers.

| Polymer | *Mn*[kDa] | $E_g^{opt}$[eV] | HOMO[eV] | LUMO[eV] | Device structure | $\mu_{h,max}$; $\mu_{e,max}$ [cm$^2$ V$^{-1}$ s$^{-1}$] | Refs |
|---|---|---|---|---|---|---|---|
| PDBPyBT | — | 1.36 | -5.69 | -4.33 | BGBC | 2.78; 6.30 | 127 |
| PDPP-4FTVT | 53 | 1.3 | -5.17 | -3.53 | BGTC | 3.40; 5.86 | 128 |
| (l-C18)-DPP-(b-C17)-BTZ | 63 | 1.25 | -5.15 | -3.9 | TGBC | 2.40; 1.50 | 49 |
| PDPP2TzBDT | — | — | — | −4.00 | BGTC | 5.47; 5.33 | 59 |
| PPyDPP1-BT | 62.1 | 1.66 | -5.65 | -3.74 | TGBC | 0.86; 1.05 | 129 |
| CNTVT:SVS | 226 | 1.29 | -5.59 | -4.30 | TGBC | 3.15; 3.03 | 130 |
| P2DPP-TT | 61.9 | 1.28 | -5.37 | -3.54 | TGBC | 4.16; 3.01 | 131 |
| P2DPP-BT | 46.9 | 1.29 | -5.44 | -3.5 | TGBC | 2.99; 2.60 | 131 |
| PDBD-Se | 64.6 | 1.18 | -5.03 | -3.85 | TGBC | 8.90; 7.71 | 132 |
| Meso-SeDPPBTz | 4.9 | 1.09 | −5.07 | −3.98 | TGBC | 2.30; 2.20 | 133 |
| PTFDFT | 24.4 | 1.25 | -5.28 | -3.79 | TGBC | 1.08; 2.23 | 134 |
| **P1FIID-2FBT** | **50.5** | **1.57** | **-5.63** | **-3.55** | **TGBC** | **6.41; 6.76** | **135** |
| **PIBDF-BTO** | **51.5** | **0.97** | **-5.00** | **-4.03** | **TGBC** | **0.27; 0.22** | **136** |
| **PBAID3-2FBT** | **40.6** | **1.27** | **-5.17** | **-3.9** | **BGTC** | **1.68; 1.37** | **137** |
| **PFIBI-BT** | **51.7** | **1.51** | **-5.01** | **-3.50** | **TGBC** | **5.50; 4.50** | **138** |
| **BDOPV-TT** | **20.9** | **1.42** | **-5.70** | **-3.88** | **TGBC** | **1.70; 1.37** | **139** |
| **PINDFBT** | **43.1** | **1.81** | **-5.65** | **-3.84** | **BGBC** | **0.51; 0.50** | **140** |

### 3.3.1 DPP-based polymers

DPP-based polymers are representative ambipolar materials with exciting charge carrier mobilities for both holes and electrons [126]. Mainly, the substitution of electron-withdrawing moieties or copolymerization of strong electron-withdrawing groups flanked by DPP units supplied effective routes for a highly balanced ambipolar polymer. A copolymer of 2-pyridinyl substituted DPP block (DBPy) and T2 (PDBPyBT) was synthesized, displaying the electron mobility up to 6.3 $cm^2 V^{-1} s^{-1}$ and the hole mobility up to 2.78 $cm^2 V^{-1} s^{-1}$, mainly owing to the reduction of steric interaction with the DPP core by 2-pyridyl substituent, resulting in a highly coplanar structure [127]. The DPP-based polymer poly[2,5-bis(2-decyltetradecyl)pyrrolo[3,4-*c*]pyrrole1,4(2*H*,5*H*)-dione-alt-5,5′-di(thiophen-2-yl)-2,2′-(*E*)-1,2-bis(3,4difluorothien-2-yl)ethene] (PDPP-4FTVT) with tetrafluoride TVT units through direct arylation polycondensation was synthesized. The incorporation of F-atoms in *β*-positions of thiophene rings not only improved the packing order of the polymer backbones, but also lowered HOMO and LUMO energy levels of the polymer, resulting in hole and electron mobilities of the OTFTs up to 3.40 and 5.86 $cm^2 V^{-1} s^{-1}$, respectively [128]. Moreover, through Suzuki polycondensation reaction, a low bandgap diketopyrrolopyrrole-benzotriazole (BTZ) copolymer ((l-$C_{18}$)-DPP-(b-$C_{17}$)-BTZ) with good solubility and highly planar conformation was synthesized, exhibiting high and near-equilibrium average electron and hole mobility of 2.4 $cm^2 V^{-1} s^{-1}$ and 1.5 $cm^2 V^{-1} s^{-1}$. This suggests that the incorporation of linear side chains on the DPP units led to an increase in thin-film order and charge-carrier mobility [49].

When thiazole was selected as the linker unit to bridge the DPP core and benzodithiophene units, the DPP-benzodithiophene-thiazole copolymer (PDPP2TzBDT) nanowire with well-ordered molecular packing adopted a distinctly "face-on" configuration molecular orientation on the surface, and exhibited a much better charge mobility of devices via *in-situ* drop-coating method ($\mu_h$ = 5.47 $cm^2 V^{-1} s^{-1}$ and $\mu_e$ = 5.33 $cm^2 V^{-1} s^{-1}$) [59]. With a similar strategy, pyridine flanked diketopyrrolopyrrole-bithiophene copolymer (PPyDPP1-BT) by copolymerizing pyridine flanked DPP (PyDPP) as the A unit and BT as the D unit possessed a highly coplanar structure, and revealed a good ambipolar transport behavior of 0.86 ($\mu_h$) and 1.05 $cm^2 V^{-1} s^{-1}$ ($\mu_e$), respectively [129].

Precise adjustion of the copolymerization ratio of the electron donor and acceptor structural unit was then employed to control the ambipolar charge transport feature of the conjugated polymer. For example, by tuning copolymerization ratio between electron transport unit poly[2,5-bis(2-octyldodecyl)pyrrolo[3,4-*c*]pyrrole-1,4(2*H*,5*H*)-dione-(*E*)-[2,2-bithiophen]-5-yl)-3-(thiophen2-yl)acrylonitrile (DPP-CNTVT) and hole transport unit DPP-SVS, a copolymer (CNTVT:SVS) with highly coplanar architecture and superior packing motif could realize the precise controllability of the transport mode of the spin-coated films, displaying well-balanced mobility values ($\mu_h$ = 3.15 cm$^2$ V$^{-1}$ s$^{-1}$; $\mu_e$ = 3.03 cm$^2$ V$^{-1}$ s$^{-1}$) [130].

In addition, DPP dimerization has been proven to be a promising strategy for ambipolar D–A polymers due to the high lackness of electrons. By using bis-DPP (diketopyrrolopyrrole dimer, 2DPP) moiety as a general acceptor to couple with common thiophene-based donor T2 and TT, two ambipolar copolymers P2DPP-T2 and P2DPP-TT with a completely coplanar structure were obtained and showed narrow bandgaps and lower LUMO energy levels, leading to high charge-carrier mobility values of $\mu_h$ and $\mu_e$ up to 2.99 and 2.60 cm$^2$ V$^{-1}$ s$^{-1}$, 4.16 and 3.01 cm$^2$ V$^{-1}$ s$^{-1}$, respectively [131]. Similarly, the introduction of T2 molecules into 2DPP selenium-based copolymers PDBD-Se via an all C-H activated cross-coupling method has recently been reported to get extremely high ambipolar mobility with $\mu_h$ and $\mu_e$ up to 8.90 and 7.71 cm$^2$ V$^{-1}$ s$^{-1}$, mainly due to the strengthened electron deficiency in 2DPP over DPP core [132].

A diketopyrrolopyrrole monomer and a benzothiadiazole derivative could react each other through the direct arylation polycondensation to form mesopolymers of number-averaged molecular weights (Mn) between 1 and 10 kDa without structural defects, low solubility and batchto-batch variation [133]. This mesopolymer meso-SeDPPBTz had a wider band gap and deeper LUMO energy level (−3.98 eV vs −3.92 eV) than the corresponding polymer poly[3,6-Bis(5-bromoselenophen-2-yl)-2,5-bis(2-octyldodecyl)pyrrolo[3,4-*c*]pyrrole-1,4(2*H*,5*H*)-dione-4,7-bis(4,4,5,5-tetramethyl-1,3,2-dioxaborolan-2-yl)benzo[*c*][1,2,5]thiadiazole] (poly-SeDPPBTz), expressing a hole mobility value of 2.3 cm$^2$ V$^{-1}$ s$^{-1}$ and an electron mobility of 2.2 cm$^2$ V$^{-1}$ s$^{-1}$. Lately, a half-fused DPP unit, in which a flanking thiophene unit was fused onto a DPP ring through a carbon-carbon double bond at the N-position, has been successfully used as an electron acceptor to prepare conjugated donor-acceptor polymer (9-(3-octadecylhenicosyl)-8-(thiophen-2-yl)-7H-pyrrolo[3,4-*a*]thieno[3,2-*g*]-indolizine7,10(9*H*)-dione) (PTFDFT) [134]. Compared with the polymer containing conventional DPP units, PTFDFT had a flatter skeleton, narrow bandgap and low LUMO energy level (−3.79 eV), which was conducive to the dense packing of polymer chains. The thin-film devices showed excellent ambipolar semiconductor characteristics under ambient conditions, with n-channel and p-channel mobilities reaching 2.23 and 1.08 cm$^2$ V$^{-1}$ s$^{-1}$, respectively (Fig. 11a). Moreover, the as-constructed PTFDFT CMOS inverters illustrated an ideal switching capability in the case of equivalent p-channel and n-channel OFETs with a good noise margin of 56%, and a high gain value of 141 at $V_{DD}$ = 100 V (Fig. 11b).

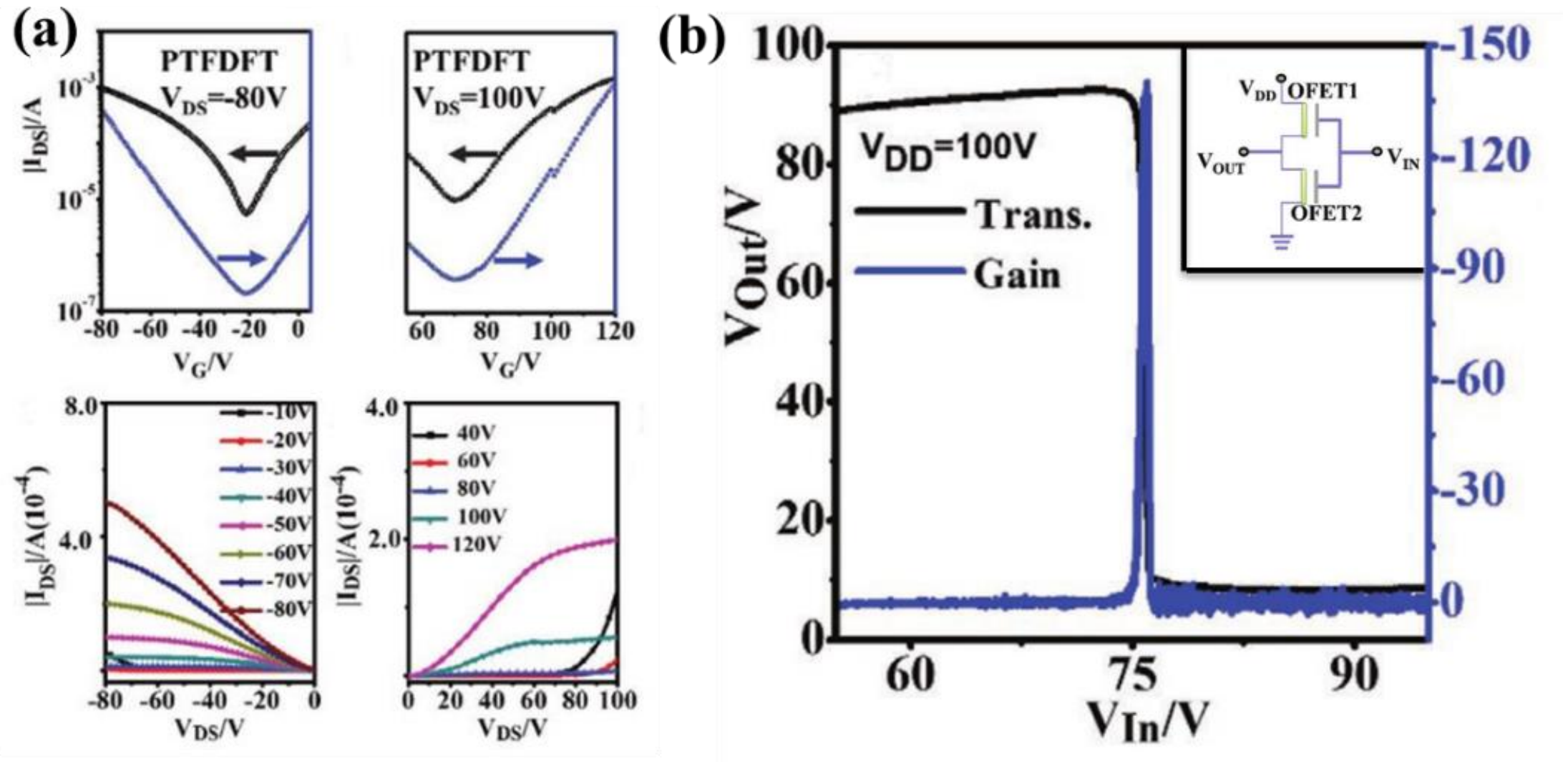

**Fig. 11** (a) Transfer and output characteristics of thermally annealed FETs with PTFDFT thin films. (b) Gain value of inverter at $V_{DD}$ = 100 V. Inset: Inverter circuit configuration. Reproduced from Ref. [134].

### 3.3.2 IID-based polymers

The energy level and main-chain conformation of IID-based polymers are mainly controlled by the number of fluorine atoms and/or the substitution position of donor/acceptor units. Through fluorination, isoindigo-3,3′-difluoro-2,2′bithiophene-based copolymer (P1FIID-2FBT) exhibited a highly-balanced $\mu_h$ of 6.41 and $\mu_e$ of 6.76 cm$^2$ V$^{-1}$ s$^{-1}$. The reason is that the planar skeleton, formed through F··S non-covalent interactions, leads to an enhanced coplanarity, lower energy level and higher crystallinity [135]. Poly(isoindigo benzodifurandione-bithiophene-alkoxyl) (PIBDF-BTO) was synthesized by introducing electron withdrawing benzodifurandione (BDF) into the acceptor isoindigo unit to deepen HOMO and LUMO energy levels, which could provide better polymer thin-film chain crystallinity and connectivity by increasing intermolecular interactions and good coplanarity, and hole and electron mobilities were calculated to be 0.27 and 0.22 cm$^2$ V$^{-1}$ s$^{-1}$, respectively [136]. A donor-acceptor (D-A) conjugated polymer poly([3*E*,8*E*]-3,8-bis(6-bromo-7-aza-2-oxoindolin-3-ylidene)6,8-dihydroindolo[7,6-*g*]indole-2,7(1*H*,3*H*)-dione-3,3′-difluoro-2,2′bithiophene) (PBAID3-2FBT) based on bis-azaisoindigo unit (BAID) was synthesized with fine level regulation through introducing nitrogen and fluorine heteroatoms and optimizing the side chain, and presented co-planar backbones, relatively low LUMO levels (-3.90 eV), and the mobilities of $\mu_h$ = 1.68 and $\mu_e$ = 1.37 cm$^2$ V$^{-1}$ s$^{-1}$ [137]. Recently, the large-area oriented poly(fluoroisoindigodifluorobithiophene-fluoroisoindigo-bithiophene) (PFIBI-BT) film was prepared by the bar-coating method through Landau–Levich process in air. The as-prepared films provided the hole mobility as high as 5.5 cm$^2$ V$^{-1}$ s$^{-1}$ and the electron mobility up to 4.5 cm$^2$ V$^{-1}$ s$^{-1}$, due to the highly ordered grains and well-aligning edge-on molecules [138].

### 3.3.3 BDOPV-based polymers

As BDOPV-based polymers exhibited electron-dominating transport due to the low LUMO energy level of the BDOPV unit, an ambipolar transport was predicted to be achieved by the introduction of extra electron-rich groups [121]. For instance, by introducing an electron-rich thieno[3,2-b]thiophene unit into the BDOPV-based polymer, 3,7-bis((E)-2-oxoindolin-3-ylidene)-3,7-dihydrobenzo[1,2-b:4,5-b']difuran-2,6-dione-thieno[3,2-b]thiophene-based polymer (BDOPV-TT) polymer was synthesized, and the balanced charge transport performance of $\mu_h$ = 1.7 cm$^2$ V$^{-1}$ s$^{-1}$ and $\mu_e$ = 1.37 cm$^2$ V$^{-1}$ s$^{-1}$ could be realized from its spin-coated thin film with the good lamellar edge-on packing [139]. Furthermore, by modifying the fused phenyl ring of the BDOPV unit to a naphthalene motif with a larger conjugated core to enhance the intermolecular *π-π* overlap between adjacent chains, poly[(3*E*,8*E*)-3,8-Bis(2-oxoindolin-3-ylidene)naphtho-[1,2-b:5,6b']difuran-2,7(3*H*,8*H*)-dione-bithiophene] PINDFBT with lower frontier energy levels was obtained to display $\mu_h$ and $\mu_e$ of OTFTs of 0.51 and 0.50 cm$^2$ V$^{-1}$ s$^{-1}$, respectively [140].

## 4. Ordered alignment of organic polymers

In recent years, because a variety of organic conjugated polymers have been developed, the field of high-performance polymer-based devices moves forward rapidly. However, how to fabricate the ideal polymer devices through controlling the alignment of organic semiconductors is still an urgent issue. Therefore, searching novel effective strategies to promote the alignment of interwoven conjugated polymer chains in the solution phase, molten state, or curing procedure is highly desirable. From the perspective of organic conjugated polymers, this part summarizes how to optimize and improve their stacking order by different processing methods to further improve the transport characteristics of carriers.

### 4.1 Aligned semiconducting polymer films

Although many preparation methods have been developed, the simplest and most convenient method is the drop-casting approach. Dropping a semiconductor solution on the substrates could form the ordered thin films after the solvent evaporation (Fig. 12a). Due to the difficulty in the uniformity controlling, there exist many vacancies in the final film, which decreases the performance of the OFET. By filling the chamber with solvent vapor, the evaporation of the solvent will be greatly slowed down, allowing enough time to form a well-oriented ordered structure, which can reduce possible defects in the semiconductor film or at the interface layers [26]. In 2006, the mobility of the P3HT film, prepared by this SVE DC method, reached $1.2 \times 10^{-1}$ cm$^2$ V$^{-1}$ s$^{-1}$, which was higher than that by the spin-coating method [27]. Similar techniques have also been used to enhance the orientation of DPP-based conjugated polymer films. For example, the donor-acceptor copolymer 3,6-bis(thiophen-2-yl)-*N,N'*-bis(2-octyl-1-dodecyl)-1,4-dioxo-pyrrolo[3,4-c]pyrrole and thieno-[3,2-b]thiophene (PDBT-TT) molecules under solvent vapor atmosphere were not affected by additional driving force during the evaporation process, allowing the molecules to slowly assemble into parallel fibers and increase the dichroic ratio of the SVE DC film from 1.43 to 4.83 (Fig. 12b) [26]. Besides, the drop-casting method involving capillarity has also been used. A sandwiched tunnel system, composing of two nano-grooved silicon substrates and a pair of surface-wetting glass spacers, was designed to realize the unidirectional polymer alignment by capillarity [78]. The CDT-PT polymers coated in this tunnel-like system tended to form oriented films with an ultrahigh hole mobility (21.3 $cm^2$ $V^{-1}$ $s^{-1}$).

Spin-coating is the most widely used technique for the preparation of OTFTs, where the solution on the substrate surface is uniformly diffused by centrifugal action. This method is convenient and simple, but the disadvantages like waste of materials, difficulty for large-area substrates, and less control of film crystallinity could not be neglected. Therefore, many studies have been conducted to optimize this technology, such as antisolvent/additive addition [99, 141]. Electron transport P(NDI2OD-T2) film, blended with electron donor tetrathi-afulvalene (TTF) or electron acceptor tetracyanoquinodimethane (TCNQ), possessed edge-on and face-on bimodal texture with long-range oriented order [142]. The conjugated small molecules, uniformly distributed in the semiconducting polymer, could be located in the defect sites and amorphous regions to establish efficient charge transport paths. Compared with neat P(NDI2OD-T2) polymer semiconductor (0.63 $cm^2$ $V^{-1}$ $s^{-1}$), P(NDI2OD-T2) OFETs containing TTF or TCNQ exhibited a higher average electron mobility of 1.36 or 1.61 $cm^2$ $V^{-1}$ $s^{-1}$.

Different from the general spin-coating method, the off-center spin-coating is to place the substrate with the polymer solution away from the center of rotation to prepare thin-films with a clear orientation under the unidirectional force (Fig. 12c) [143-145]. For P3HT dissolved in the marginal solvent, the off-center spin-coating provided an external force to form highly-crystalline, aligned, and ordered nanowires in the films [42]. The mobility of OFETs based on highly oriented and crystalline P3HT nanowires (0.053 $cm^2$ $V^{-1}$ $s^{-1}$) was 50 times higher than the spin-coated devices prepared by chloroform solvent. Combining the off-center spin-coating and kinetically controlled crystallization, the P(NDI2OD-T2) film dried at 100°C showed aligned polymer packing structures (Fig. 12d), with the mobility (3.99 $cm^2$ $V^{-1}$ $s^{-1}$) in two orders of magnitude larger than that of the film prepared through naturally drying at 25°C [94].

The dip-coating method can strengthen the macroscopic order of the film by controlling the evaporation rate of the solvent and the coating speed (Fig. 12e) [146, 147]. Two important factors affect the quality of as-obtained film: solvent evaporation and viscous effects [148]. When the volatilization rate of the solvent at the meniscus, where the solution contacting the substrate is balanced with the rate of compound deposition, an ordered thin film can be formed [149]. Highly oriented thin films of the IID-based conjugated polymer, poly-[1,1'-bis(4-decyltetradecyl)-6-methyl-6'-(5'-methyl-[2,2'-bithiophen]-5-yl)-[3,3'-biindolinylidene]-2,2'-dione] (PII-2T), was prepared by adjusting the concentration and the pulling rate [29]. These aligned conjugated films with a bimodal distribution of edge-on and face-on motifs (Fig. 12f) showed enhanced charge carrier mobilities of 8.3 $cm^2$ $V^{-1}$ $s^{-1}$ in the saturation region and 6.3

cm$^2$ V$^{-1}$ s$^{-1}$ in the linear region, ten times higher than the mobility of spin-coated films. Compared with some cumbersome or inefficient methods, dip-coating with the characteristics of scalable processing and high material utilization is a simple and versatile way to achieve industrial production.

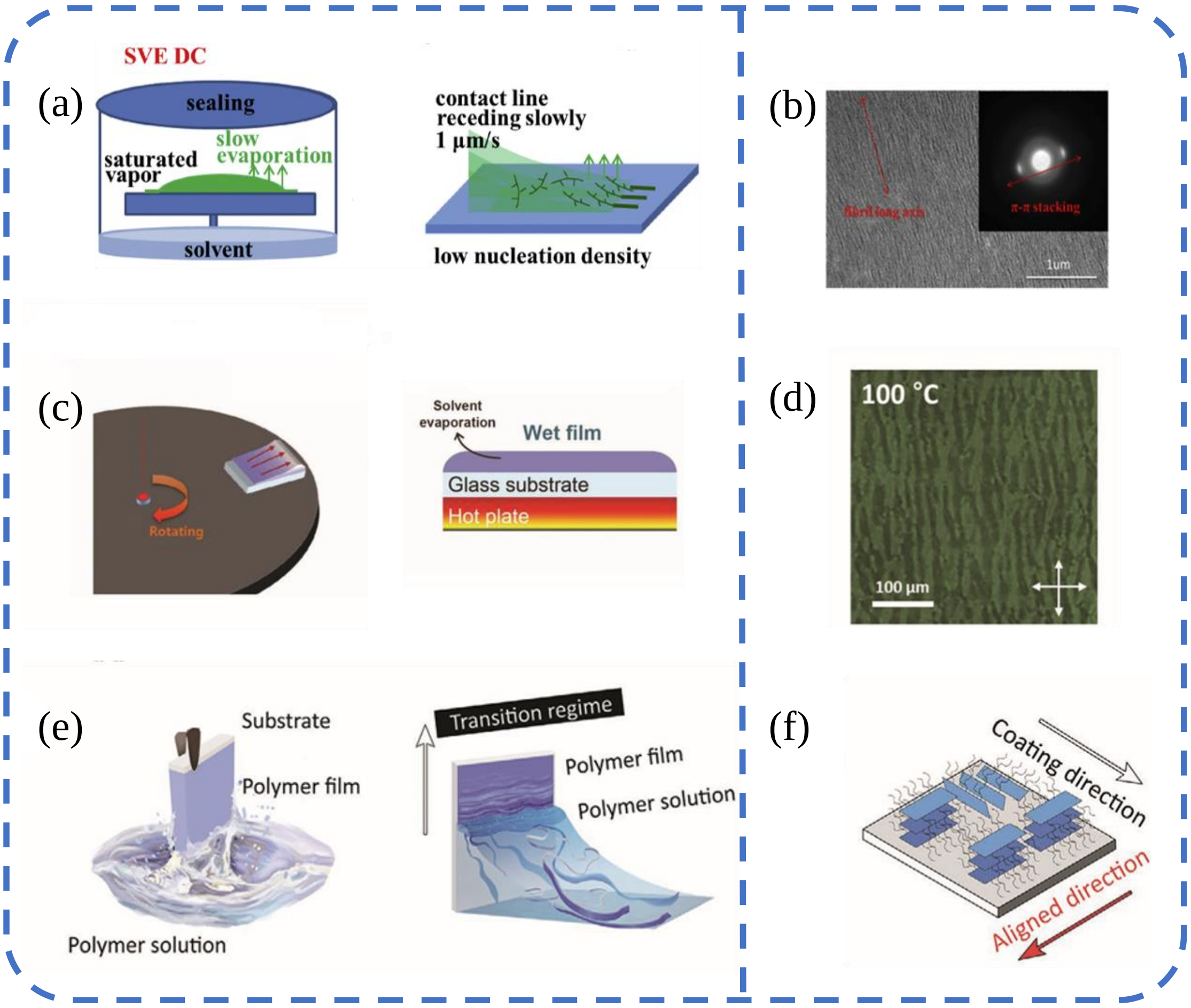

**Fig. 12** (a) The schematic illustration of the SVE DC and the process of contact line receding with evaporation. (b) TEM image of the SVE DC film. The inset shows the selected area electron diffraction pattern of the film. Reproduced from Ref. [26]. (c) The illustration of off-center spin-coating and kinetically controlled KCC processes. (d) Polarized optical microscopy images of P(NDI2OD-T2) thin films formed at 100 °C. Reproduced from Ref. [94]. (e) Schematic illustrations of the dip-coating process to deposit polymer films and transition regimes. (f) Polymer chains with edge-on and face-on orientation after dip-coating. Reproduced from Ref. [29].

Since the economical solution-deposition method is the key to the manufacturing of next generation electronics such as low-cost OFETs, shear printing has attracted widespread attention [28, 150]. However, the operation factors including solution concentration, substrate temperature, scraper angle, distance between the scraper and substrate, and shearing speed are not easy to be controlled. For highly stretchable films of poly-[2,5-bis(7-decylnonadecyl)pyrrolo[3,4-c]pyrrole-1,4-(2 H,5 H)-dione-(E)-(1,2-bis(5-(thiophen-2-yl)selenophen-2-yl)ethene) (DPPDTSE) and SEBS sheared by a blade with microgrooves, the $\pi$–$\pi$ stacking direction was perpendicular to the alignment direction of nanofibres [30]. The nanoscale spatial confinement aligned chain conformation and promoted short-range $\pi$–$\pi$ ordering to reduce the energetic barrier for charge carrier transport. As a result, the calculated linear mobility (0.9 $cm^2$ $V^{-1}$ $s^{-1}$ ) of the solution shearing film was about twice times higher than that of the spin-coated film (0.5 $cm^2$ $V^{-1}$ $s^{-1}$ ).

A reasonable brush design is also introduced to influence the orientation of the polymer thin-film (Fig. 13a). The oriented squamae along the natural hair axis enhances the backbone alignment, leading to better carrier transport abilities for many polymers including P3HT, P(NDI2OD-T2), poly[4,8-bis(5-(2-ethylhexyl)thiophen-2-yl)benzo-[1,2-b;4,5-b′]dithiophene-2,6-diyl-alt-(4-(2-ethylhexyl)-3-fluorothieno[3,4-b]thiophene-)-2-carboxylate-2–6-diyl)] (PBDTT-FTTE), and the low-crystallinity conducting poly(3,4-ethylenedioxythiophene):poly(styrenesulfonate) (PEDOT:PSS) [151]. The shearing process promotes the backbone alignment, thus facilitates aggregation and enhances charge transport mobility, but cannot necessarily induce charge transport anisotropy. Herein, the edge-on orientation of P3HT molecules benefited to good intrachain and interchain transport, and the mobilities (whether parallel (0.18 ± 0.03 $cm^2$ $V^{-1}$ $s^{-1}$) or perpendicular (0.15 ± 0.02 $cm^2$ $V^{-1}$ $s^{-1}$) to the polymer main chain) were basically similar. Different from the orientation of P3HT (Fig. 13b), the face-on molecular orientation of P(NDI2OD-T2) couldnot provide a good channel for interchain transport (Fig. 13c), so the mobility along the polymer chain (2.0 ± 0.33 $cm^2$ $V^{-1}$ $s^{-1}$) was ten times higher than that in the vertical direction (0.24 ± 0.12 $cm^2$ $V^{-1}$ $s^{-1}$).

Bar-coating, similar to shear printing, is an oriented film preparation with different shearing tools. A stainless-steel bar is surrounded by different types of steel wires to govern the film thickness through the gap (Fig 13d) [138]. During this process, highly ordered films can be achieved by optimizing the diameters of the steel wires, solution concentration, bar moving speed, and substrate temperature. For example, the PFIBI-BT films with edge-on orientation showed hole and electron mobility up to 5.5 $cm^2$ $V^{-1}$ $s^{-1}$ and 4.5 $cm^2$ $V^{-1}$ $s^{-1}$, which are nine times higher than those in the films prepared by spin-coating process. Besides, bi-functional semiconductor–dielectric layers for OFETs were fabricated by bar-coating of P3HT and PMMA mixed solution in one-step deposition [152]. Because of the well-ordered molecular arrangement, the maximum and average value of field-effect mobility calculated from the transfer characteristics in the saturation regime were 0.02 and 0.005 $cm^2$ $V^{-1}$ $s^{-1}$, respectively.

However, the previously-reported electrical properties of several conjugated polymers do not show a sensitive positive correlation with the degree of backbone alignment. For example, the mobility of the DPP-C3 films with excellent crystallinity was only 0.015 $cm^2$ $V^{-1}$ $s^{-1}$, which was much lower than that of the blending film (1.14 $cm^2$ $V^{-1}$ $s^{-1}$) containing 1 wt% DPP-C0 with poor crystallinity. Besides, the shear deposition of highly conductive PEDOT:PSS films yielded a record high conductivity 4,600 vs. 860 S $cm^{-1}$ for the spin-coated films, but transport anisotropy and optical dichroic ratio (1.6) were negligible [153]. Thus, the interplay between thin-film microstructure evolution and charge transport is not very clear, and the film preparation process still has a lot of room for development in exploring the relationship between performance and structure.

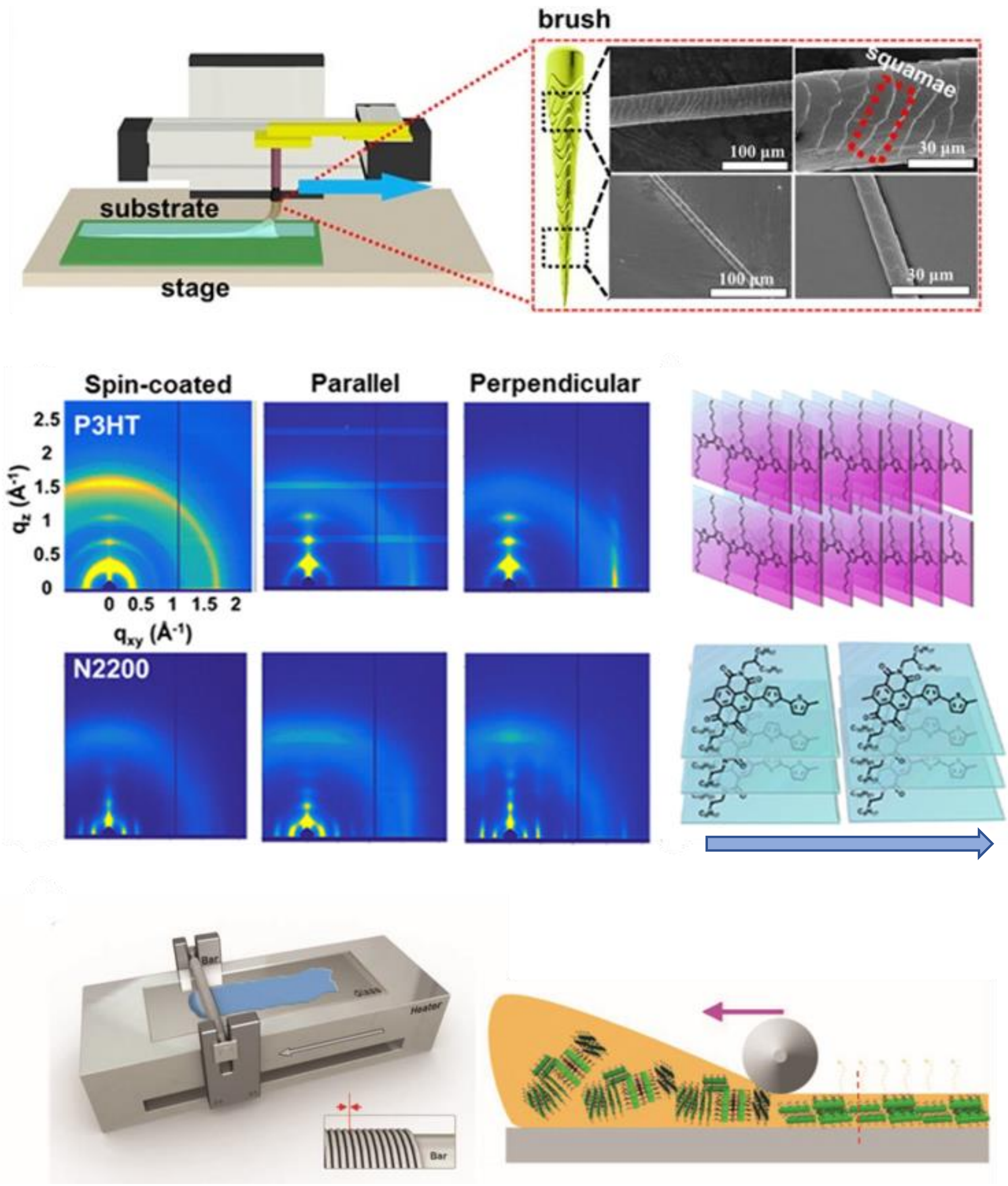

**Fig. 13** (a) Instrument schematic diagram of shear printing setup and SEM images of the brush with squamae structures. GIWAXS patterns of (b) P3HT and (c) P(NDI2OD-T2). The rightmost models correspond to the molecular chain stacking of the P3HT and P(NDI2OD-T2) films. Reproduced from Ref. [151]. (d) Instrument schematic diagram of the thin film processing by bar-coating and schematic diagram of polymer molecules arranged by edge-on. Reproduced from Ref. [138].

## 4.2 Ordered alignment of polymer single crystal/nanowire

Although the oriented polymer film contains reduced defects, it is still difficult to exert the intrinsic properties of the polymers. Thus, the preparations of defect-free conjugated polymer single-crystal devices have attracted a lot of interests. The polymers need to cross excessively large free energy barrier to get a highly-ordered state, which is hard for them to self-assemble from solution or molten state to a single crystal or high crystallinity nanostructure. Most of the polymer single crystals reported so far are grown by solution assembly methods including solvent-assisted crystallization, mixed solvent self-assembly, solution annealing, self-seeding, etc.

Initial reports on solution assembly were mainly based on P3ATs. For example, in 1993, the earliest P3HT nanowires were grown from the diluted cyclohexanone solution cooling and the mobility of a regioregular P3HT single nanowire is 0.02-0.06 $cm^2\ V^{-1}\ s^{-1}$ [33]. Molecular weight, crystallization temperature, and solution concentration affect the crystallization behavior in solution. When the molecular weight exceeds the critical value, two-dimensional P3HT nanorods can be obtained by slow crystallization growth and the hole mobility can reach 0.012 $cm^2\ V^{-1}\ s^{-1}$ [154]. Single crystals of conjugated polymer were prepared for the first time in 2006 by the solvent-assisted crystallization method in a confined space [155]. The selected area electron diffraction analysis showed that the long axis direction of the P3HT crystal was along the direction of $\pi$-$\pi$ stacking. By constantly adjusting the solvent type and solution concentration, the single nanowire of 1,2-bis(5-(thiophen-2-yl)selenophen-2-yl)ethene-DPP (DPPBTSPE), DDP polymer based on selenophene, showed the highest mobility up to 24 $cm^2\ V^{-1}\ s^{-1}$, which was nearly 20 times higher than that of the corresponding thin-film devices [156]. Besides, by changing the aromatic heterocycles connected to DPP (thiophene ring or thiazole ring), the stacking way of polymer chains in the nanowire can be regulated [59]. Structural analysis showed that in the thiophene-bridged PDPP2TBDT nanowires, the molecular chains were arranged in an edge-on manner (Fig. 14a), while the thiazole-bridged polymer PDPP2TzBD stacked in a face-on manner (Fig. 14b). Because the nanowires grew along the polymer chain, this well-ordered molecular stacking contributed to intrachain charge transport and thus higher carrier mobility. PDPP2TBDT nanowires presented hole-dominating mobility of 7.42 $cm^2\ V^{-1}\ s^{-1}$ and poor electron transport behavior of 0.0034 $cm^2\ V^{-1}\ s^{-1}$ (Figs. 14(c) and (d)), in the meanwhile, the PDPP2TzBD nanowires had a relatively balanced bipolarity with the hole and electron mobilities of 5.47 $cm^2\ V^{-1}\ s^{-1}$ and 5.33 $cm^2\ V^{-1}\ s^{-1}$, respectively (Figs. 14(e) and (f)).

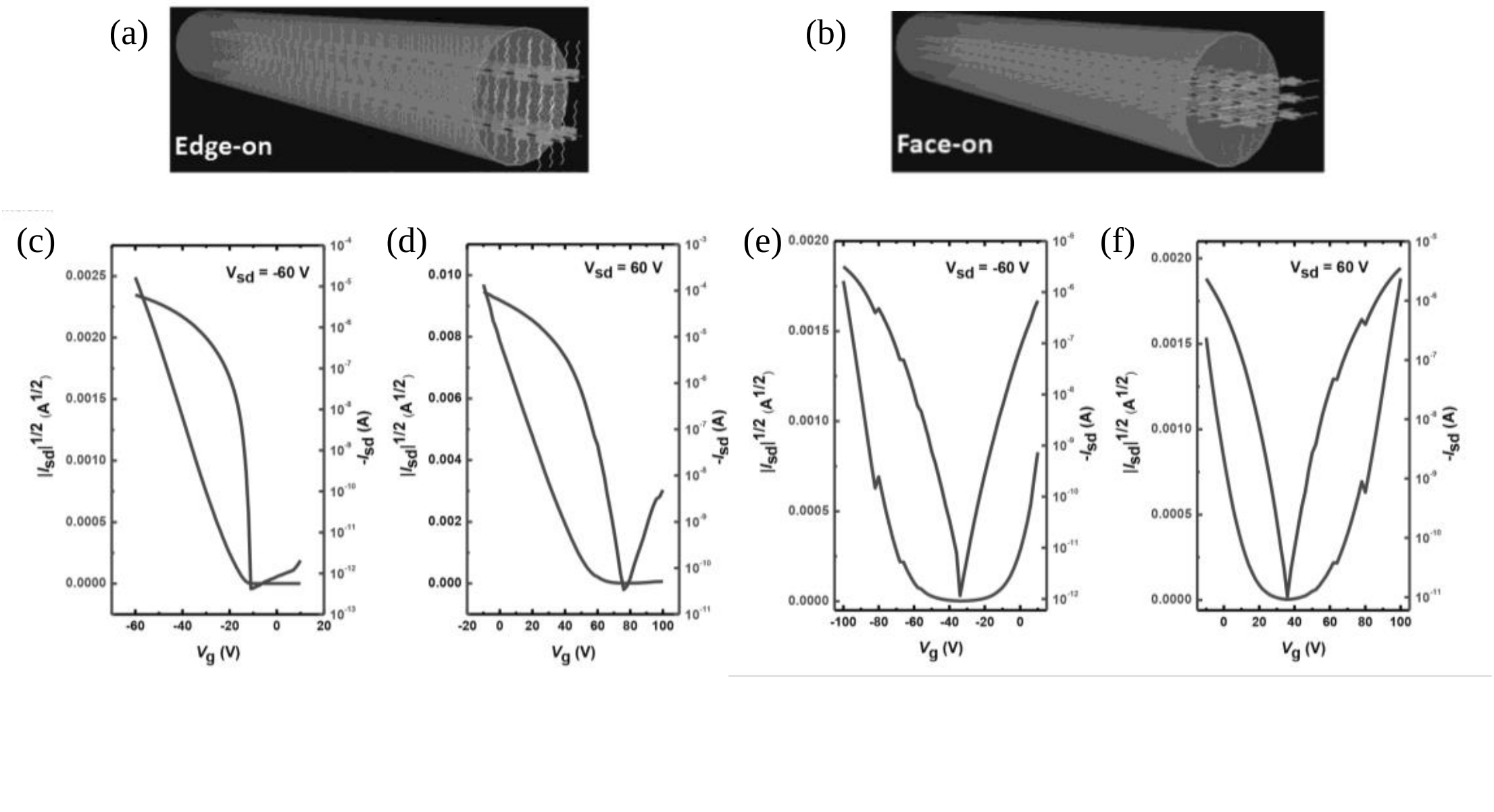

**Fig. 14** (a) Schematic diagram of PDPP2TBDT and PDPP2Tz-BDT nanowire molecular packing in the 1D nanocrystals. (b) Schematic diagram of the effect of different main solvents and cosolvents. (a, c) p-type and (b, d) n-type transfer characteristics for the (a, b) PDPP2TBDT and (c, d) PDPP2TzBDT FETs with top contact configuration. Reproduced from Ref. [59].

The mixed solvent self-assembly refers to the procedure that a solution containing well-dissolved polymers is quickly injected into a large amount of poor solvents. Because of the reduced solubility in poor solvents, the polymers firstly reach to a supersaturation state, then gradually nucleate and finally precipitate into crystals. During the assembly process, the polarity, ratio, and concentration of the solvent will largely affect the nucleation and the growth rate of polymer crystal phase. Thus, choosing a suitable solvent system and the mixed ratio becomes very important. The nucleation process of 3,6-bis-(thiophen-2-yl)-*N,N'*-bis(2-decyl-1-tetradecyl)-1,4-dioxo-pyrrolo[3,4-c]pyrrole and thieno[3,2-b]thiophene (DT-PDBT-TT) nanowires can be controlled by tuning marginal cosolvent ratio (Fig. 15a) [34]. As the main solvent (trichloro ethylene) evaporated, the core or polymer chains tended to aggregate with each other, whereas the appropriate marginal cosolvent (o-dichlorobenzene : anisole = 1:1) prevented the aggregation and provided the space for the growth of highly crystalline nanowires.

The solvent-vapor annealing technique is basically a re-self-assembly process under solvent vapor atmosphere. The solvent can assist the immigration of the polymer segments so that the molecules undergo secondary self-assembly to improve the nanostructure crystallinity. This solvent-vapor annealing method was applied to the preparation of polythiophene single crystals, in which the poly(3-octylthiophene) (P3OT) molecules partially dissolved and self-assembled into needle-shaped micro-nano crystals with regular molecular arrangement [157]. Structural analysis showed that P3OT molecules are stacked and arranged in such a way that the side chain is perpendicular to the substrate and the main chain of the molecular backbone was along the long axis of the nanowire. Later, the same method was used to further successfully prepare P3HT micro-nanocrystals [158]. Perhaps due to the influence of too many defects in the crystal or residual solvent, the single crystal nanowire had a low mobility. Recent studies have shown that during the solvent annealing process, when the solubility parameters of the solvent and polymers are similar to each other, it is more likely to form a regularly-arranged aggregate structure of the polymer [159].

The self-seeding method is one of the most common ways in traditional polymer crystallization. In the solution, there might still exist some polymer crystals even most of polymers are dissolved. Thus, these undissolved crystals could act as nuclei when the temperature decreases. Unlike common heterogeneous nucleation, the resulting crystal grown from self-seeding is pure, hence self-seeding is a successful method for growing polymer single crystals [35]. When P3HT solution is heated to the seed temperature (TSS), lower than the total dissolution temperature, and then cooled to the crystallization temperature (TC), there are some undissolved crystals in the solution acting as nuclei (Fig. 15b) [36]. Higher TSS leads to fewer nuclei, leading to larger and fewer high-quality single crystals. The solar cell, prepared from the solution through the refrigeration and heating cycle method, has a higher solar energy conversion efficiency than the conventional thermal annealing cell. Afterwards, the researchers realized the precise control of the length of the P3HT fibers by adjusting the TSS to yield fibres with relatively low-length dispersities [160]. The advantage of the self-seeding method is that not only the size and quality of the micro-nano crystals can be fully regulated by controlling the temperature, solubility, and polymer concentration of the solution, but also the crystallinity of the molecules without additional treatment can be increased.

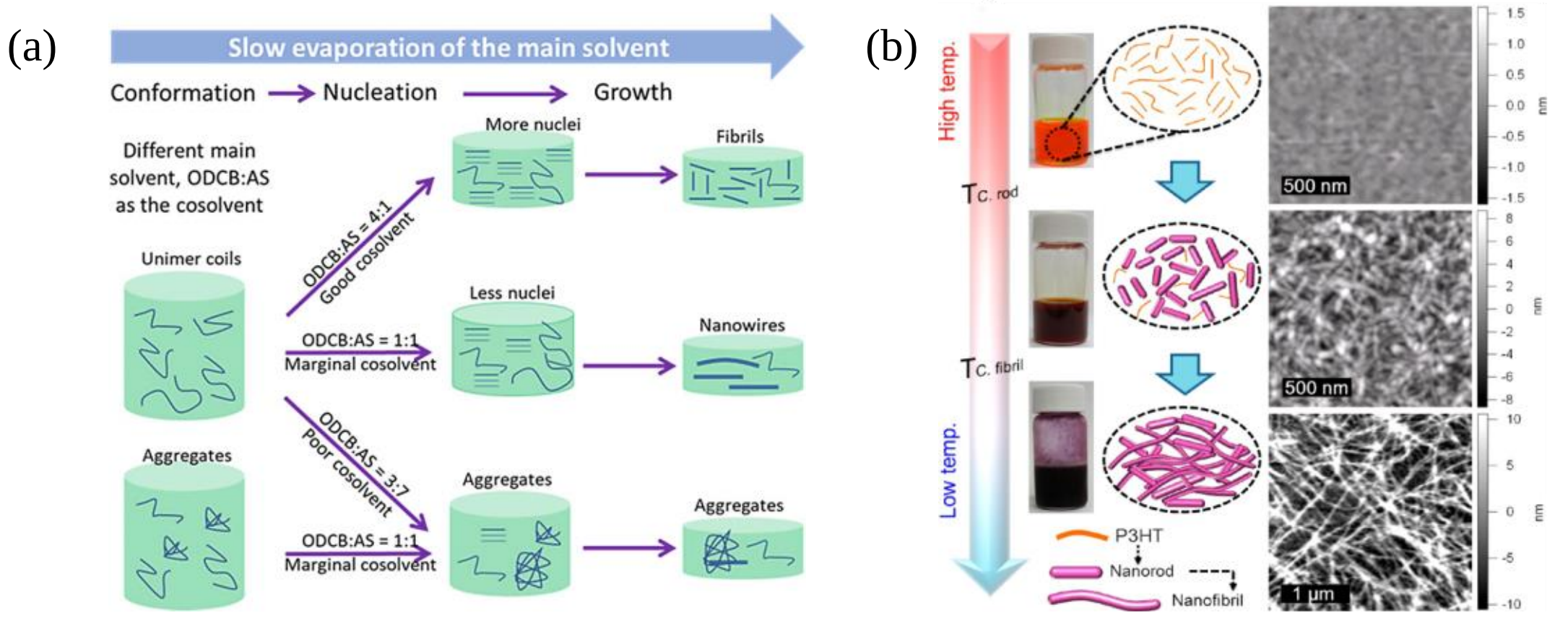

**Fig. 15** (a) Schematic diagram of the effect of different main solvents and cosolvents. Reproduced from Ref. [34]. (b) Schematic illustration of the transition of P3HT conformation from the dissolved polymer chains to the nanorods and the nanofibrils as the temperature of the solution was decreased. The digital images and the corresponding Atomic Force Microscope images were taken from an m-xylene solution (0.25 wt %). The color of the solution changed from orange to light purple, and then to deep purple. Reproduced from Ref. [36].

## 5. Conclusion

Nowadays, the charge-carrier transporting performance exceeding that of amorphous silicon is required to meet the demands of the practical applications in flexible electronic devices. During the past two decades, the development of high-performance semiconducting polymers has made great progress. Especially for p-type polymer, ultra-high hole mobility exceeding 10 $cm^2 V^{-1} s^{-1}$ has been realized. For the relatively rare and much sensitive ambipolar/n-type polymers, several chemical modification strategies have been employed, including heteroatom substitution, side-chain engineering, and novel electron-deficient unit incorporation, which in turn helps us to understand the structure-property relationships of conjugated polymers. Besides, various manufacturing techniques have been used to regulate the aggregate structures of conjugated polymers for efficient charge carrier pathway, since the macromolecular $\pi$-conjugated chains are very long and easily entangled with each other. Creating a strong solvent atmosphere, heating to a suitable assembly temperature, and applying a one-way external force into the nanostructure growth process are proven to be powerful tools to promote the orderly accumulation of polymer chains. Among them, solution shearing and bar-coating as promising manufacturing technologies have the advantages of high efficiency, material saving, low cost, and easy manufacturing on large-area substrates. Although the progress is exciting, there still exist some critical issues to be overcomed before actual commercialization. For example, how to achieve mass production of semiconducting polymers is still a huge challenge. Polymers acting as the most promising candidates for flexible and stretchable electronics with biological compatibility need deep exploitation. Overall, the realization of high-performance flexible polymer OFET devices that can be handled by facile ways is very attractive and important to bring a bright future to organic electronics.

**Acknowledgements** This work was supported by the National Natural Science Foundation of China (21602113 and 61774087), 1311 Research Foundation of Nanjing University of Posts & Telecommunications and Jiangsu Specially Appointed Professor Foundation. QZ thanks the financial support from City University of Hong Hong, State Key Laboratory of Supramolecular Structure and Materials, Jilin University (sklssm2020041).

## TOC

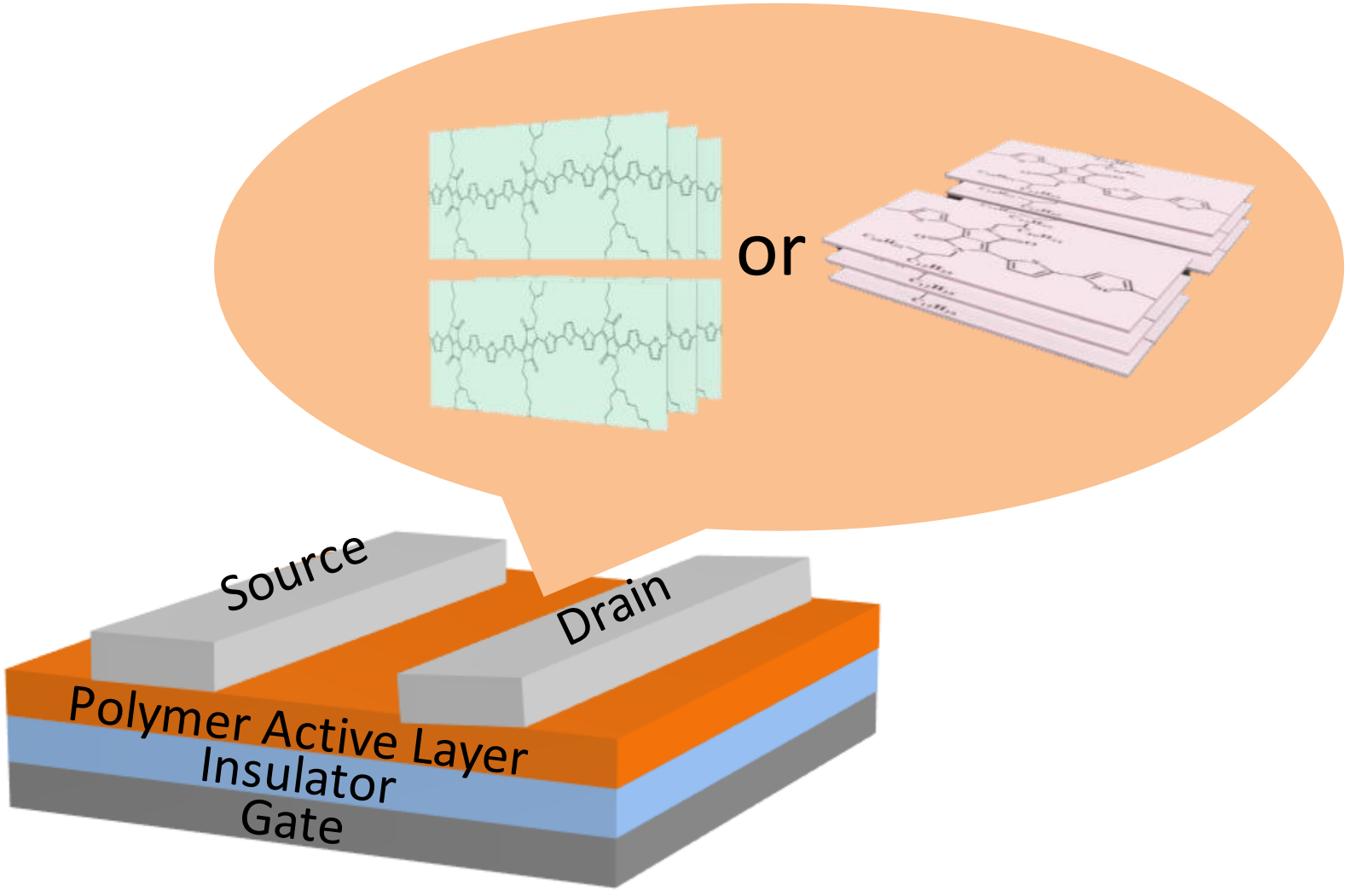